\documentclass[11pt]{article}

\usepackage{graphicx}
\usepackage{amssymb}
\usepackage{epstopdf}
\usepackage{url}
\usepackage{amsmath}
\usepackage{mathtools}
\usepackage{geometry}
\usepackage{booktabs}
\usepackage[format=hang,labelfont=bf]{caption}
\usepackage{float}
\usepackage[parfill]{parskip}
\usepackage{tensor}
\usepackage{natbib}
\usepackage[inline]{enumitem}
\usepackage{color}

\usepackage[titletoc,toc,title]{appendix}
\usepackage{chngcntr}

\usepackage[parfill]{parskip}

\usepackage{authblk}

\usepackage[format=hang,labelfont=bf]{caption}
\usepackage[raggedright]{titlesec}

\usepackage{array}
\newcolumntype{L}[1]{>{\raggedright\let\newline\\\arraybackslash\hspace{0pt}}m{#1}}
\newcolumntype{C}[1]{>{\centering\let\newline\\\arraybackslash\hspace{0pt}}m{#1}}
\newcolumntype{R}[1]{>{\raggedleft\let\newline\\\arraybackslash\hspace{0pt}}m{#1}}

\usepackage[table]{xcolor}
\definecolor{lightgray}{gray}{0.95}

\newcommand{\mbf}{\mathbf}
\newcommand{\qf}{\tensor*[_5]{\mbox{q}}{_0}}
\newcommand{\qoz}{\tensor*[_1]{\mbox{q}}{_0}}
\newcommand{\qoo}{\tensor*[_1]{\mbox{q}}{_1}}

\newcommand{\qof}{\tensor*[_1]{\mbox{q}}{_4}}
\newcommand{\qff}{\tensor*[_{45}]{\mbox{q}}{_{15}}}
\newcommand{\qox}{\tensor*[_{1}]{\mbox{q}}{_{x}}}
\newcommand{\qoa}{\tensor*[_{1}]{\mbox{q}}{_{a}}}
\newcommand{\qfx}{\tensor*[_{5}]{\mbox{q}}{_{x}}}
\newcommand{\qnx}{\tensor*[_{n}]{\mbox{q}}{_{x}}}
\newcommand{\ez}{{\mbox{e}}{_0}}
\newcommand{\mfx}{\tensor*[_{5}]{\mbox{m}}{_{x}}}

\newcommand{\logit}{\mbox{logit}}
\newcommand{\expit}{\mbox{expit}}

\makeatletter
\newcommand\Label[1]{&\refstepcounter{equation}(\theequation)\ltx@label{#1}&}
\makeatother

\usepackage{auto-pst-pdf}
\usepackage{breakcites}

\title{A General Age-Specific Mortality Model with An Example Indexed by Child or Child/Adult Mortality }

\author[1,2,3,4,*]{Samuel J. Clark}

\affil[1]{Department of Sociology, The Ohio State University}
\affil[2]{MRC/Wits Rural Public Health and Health Transitions Research Unit (Agincourt), School of Public Health, Faculty of Health Sciences, University of the Witwatersrand}
\affil[3]{The ALPHA Network, London School of Hygiene and Tropical Medicine, London, UK}
\affil[4]{The INDEPTH Network, Accra, Ghana}
\affil[*] {Contact: work@samclark.net, 206.303.9620}

\date{\today \vspace{-1.5cm}}                                           

\begin{document}
\sloppy

\pagenumbering{roman} %

\maketitle
\thispagestyle{empty}

\vspace{2cm}
\begin{abstract}

\noindent \textbf{Background.}  The majority of countries in Africa and nearly one third of all countries require mortality models to infer complete age schedules of mortality that are required to conduct population estimates, projections/forecasts and many other tasks in demography and epidemiology.  Models that relate child mortality to mortality at other ages are particularly important because almost all countries have measures of child mortality.\\

\noindent \textbf{Objective.} 
\begin{enumerate*}[label=\arabic*)]
\item Define a general model for age-specific mortality that provides a standard way to relate covariates to age-specific mortality.
\item Calibrate that model using the relationship between child or child/adult mortality and mortality at other ages embodied in a large collection of high quality observed mortality schedules.
\item Validate the calibrated model and compare its performance to existing models.
\item Provide open source software that implements the model.
\end{enumerate*}\\

\noindent \textbf{Methods.} A general, parametrizable component model of mortality is defined using the singular value decomposition (SVD-Comp) and calibrated to the relationship between child or child/adult mortality and mortality at other ages in the observed mortality schedules of the Human Mortality Database.  Cross validation is used to validate the model, and the predictive performance of the model is compared to that of the Log-Quad model, designed to do the same thing.\\

\noindent \textbf{Results.}  Prediction and cross validation tests indicate that the child mortality-calibrated SVD-Comp is able to accurately represent the observed mortality schedules in the Human Mortality Database, is robust to the selection of mortality schedules used to calibrate it, and performs better than the Log-Quad Model.\\

\noindent \textbf{Conclusions.} The child mortality-calibrated SVD-Comp is a useful tool that can be used where child mortality is available but mortality at other ages is unknown.  Together with earlier work on an HIV prevalence-calibrated version of SVD-Comp, this work suggests that this approach is truly general and could be used to develop a wide range of additional useful models.

\end{abstract}

\newpage
\tableofcontents

\clearpage 
\pagenumbering{arabic} 

\newpage

\section{Introduction}

Complete age-specific mortality schedules are necessary inputs to a wide variety of formal demographic and epidemiological methods.  A key example is the biennial World Population Prospects (WPP) \citep{un2015} produced by the UN Population Division.  These are considered the gold standard population indicators and are used widely by other domestic and international agencies as inputs to estimation and modeling exercises.  The WPP contains estimates of time-sex-age-specific mortality, fertility and population size from 1950 to the present and forecasts of the same quantities to 2100 for all countries of the world.  Consequently each WPP update must contain full age-specific mortality schedules covering the period 1950--2100.

\begin{table}[htp]
\captionsetup{format=plain,font=normalsize,margin=0.7cm,justification=justified}
\caption{\textbf{Countries or regions with no information on either child or adult mortality.} UN countries and regions that do not have information on either child or adult mortality for the 2015 update of the World Population Prospects, with population and fraction of total population for which information is missing. \textit{Reference:} \cite{unWPPMethods2015} tables I.1b (p 5) and I.1c (p 6).}
\begin{center}
\begin{tabular}{llllllll}

\toprule
 & \multicolumn{3}{c}{Child Mortality} & &  \multicolumn{3}{c}{Adult Mortality} \\
 \cmidrule{2-4} \cmidrule{6-8}
 & Regions & \begin{tabular}{@{}l} Population \\ (millions) \end{tabular} & \begin{tabular}{@{}l} Percent \\ Population \end{tabular} & & Regions & \begin{tabular}{@{}l} Population \\ (millions) \end{tabular} &  \begin{tabular}{@{}l} Percent \\ Population \end{tabular} \\
\midrule
World & 1 & 1 & 0.0\% & & 50 & 973 & 13.2\% \\
Africa & 1 & 1 & 0.0\% & & 33 & 666 & 56.1\% \\

\bottomrule
\end{tabular}
\end{center}
\label{tab:mxMiss}
\end{table}%

Some countries in the developing world, particularly in Africa, do not yet have civil registration and vital statistic systems that function well enough to accurately report on either fertility or mortality.  Focusing on mortality, Table \ref{tab:mxMiss} displays the number of countries or world regions for which there is no information on either child mortality or adult mortality, with Africa broken out.  Because of the exhaustive coverage of household surveys investigating fertility and maternal/child health, essentially the whole world has at least some recent information on child mortality \citep{li2015wppLt}.  In contrast 50 countries around the world with a total population of nearly 1B people have no information on adult mortality, with the bulk of those in Africa -- 33 countries with a total population of 666M people. 

Mortality models are used to solve this problem and produce full age schedules of mortality.  Table \ref{tab:mxMods} describes the number of countries or world regions for which the UN Population Division must use mortality models of some kind to produce either estimates of life expectancy at birth $\ez$ or full age schedules of mortality.  Most African countries require mortality models for both, and globally 38.6\% countries require a model for $\ez$ and 32.6\% for age-specific mortality.    

\begin{table}[htp]
\captionsetup{format=plain,font=normalsize,margin=2.4cm,justification=justified}
\caption{\textbf{Countries and regions where mortality models are necessary to estimate life expectancy at birth ($\ez$) or age-specific mortality rates (ASMR).} 
Counts of the number of UN countries  and regions where mortality models were used to generate estimates of $\ez$ or age-specific mortality rates for the 2015 update of the World Population Prospects. \textit{Reference:} \cite{unWPP2015Meta}.  }
\begin{center}
\begin{tabular}{lllllll}

\toprule
&  & \multicolumn{2}{c}{$\ez$} & & \multicolumn{2}{c}{ASMR}  \\
\cmidrule{3-4} \cmidrule{6-7}
& Countries/Regions & Count & Percent & & Count & Percent \\ 
\midrule
World & 233 & 90 & 38.6\% & & 76 & 32.6\% \\
Africa & 58 & 50 & 86.2\% & & 50 & 86.2\% \\

\bottomrule
\end{tabular}
\end{center}
\label{tab:mxMods}
\end{table}

The standard approach to generating complete age schedules of mortality for countries and regions with insufficient data is to take advantage of the fact that they do have information on child mortality.  Typically, model life tables are used to extrapolate full mortality schedules from $\qf$.  The UN Population Division uses the Log Quadratic (Log-Quad) model created by \cite{wilmoth2012flexible} and updated by \cite{li2015wppLt} to do this for the many countries and world regions with incomplete mortality data, and the Institute for Health Metrics and Evaluation (IHME) uses variations on the Modifed Logit (Mod-Logit) model \citep{murray2003} to do the same.  

The commonly used model life table systems -- Regional Model Life Tables and Stable Populations \citep{coale1966}, Life Tables for Developing Countries \citep{united1982model}, Modified Logit Life Table System (Mod-Logit) \citep{murray2003, Wang2013} and Flexible Two Dimensional Mortality Model (Log-Quad) \citep{wilmoth2012flexible} -- combine a specific model structure and defined variable parameters with a set of fixed parameters that summarize the relationships between mortality at different ages in a set of observed life tables.  All are \emph{empirical} models in the sense that they summarize observed mortality and use that summary to produce arbitrary mortality schedules that are consistent with observed mortality.  They come in both regional and continuous forms.  The regional models identify and replicate commonly observed mortality patterns associated with geographic regions (and \textit{de facto} time periods) while the continuous models generate mortality patterns that vary smoothly.  The input parameters range from region and life expectancy to child and adult mortality.  

\cite{murray2003} enumerate three required characteristics of mortality models: 
\begin{enumerate*}[label=\arabic*)]
\item simplicity and ease of use,
\item comprehensive representation of the true variability in sex-age-specific mortality observed in real populations, and
\item validity that is well quantified by comparing age schedules of mortality predicted by the model to corresponding observed life tables
\end{enumerate*}.
To those I would add:
\begin{enumerate*}[label=\arabic*)]
\item generality with respect to the underlying model structure, 
\item flexibility in terms of input parameters, and 
\item an ability to handle arbitrary, including very fine-grained, age groups without having to fundamentally alter the structure of the model
\end{enumerate*}. 

This work defines a mortality modeling framework that satisfies all of those requirements, and I use it to create a mortality model that predicts single-year of age mortality schedules from $\qf$ or both $(\qf, \, \qff)$, similar to both the Mod-Logit and Log-Quad models.   The resulting model can be used to produce single year of age mortality schedules from $\qf$ alone that are consistent with observed mortality schedules, and this could be useful for those like the UN Population Division who must manipulate full age schedules of mortality but only have observed values for $\qf$.

The remainder of this article
\begin{enumerate*}[label=\arabic*)]
\item reviews existing mortality models with an emphasis on those that use a dimension-reduction approach, 
\item identifies and describes the empirical life tables used to develop the model,
\item develops and calibrates the model so that it reflects observed mortality across a wide range of settings and times,
\item uses a cross-validation approach to validate the model, and
\item compares the performance of the model with that of the Log-Quad model
\end{enumerate*}.

\section{Mortality Models}

Traditional model life tables \citep[e.g.][]{united1955age, ledermann1969nouvelles, coale1966, united1982model, murray2003, wilmoth2012flexible, Wang2013} take an inductive, empirically-driven approach to identify and parsimoniously express the regularity of mortality with age based on observed relationships in large collections of high quality life tables.  Some fertility models \citep[e.g.][]{coale1974model, lee1993modeling} do the same.  An alternative, sometimes deductive approach, can be found in the wide variety of  parametric or functional-form mortality models \cite[e.g.][]{gompertz1825nature, makeham1860law, heligman1980, li2009vitality} that define age-specific measures of mortality in an analytical form, sometimes with interpretable parameters. \cite{brass1971scale} developed an innovative new approach with his two-parameter `relational' model that has been extended and refined in many ways, \citep[for example][]{zaba1979four, murray2003}.  More recently the Log-Quad model of \cite{wilmoth2012flexible} combines empirical and functional-form approaches to mortality models.

Population forecasting has focused efforts to develop highly parsimonious, empirically-driven models of demographic age schedules. Forecasting generates many iterations of both age-specific mortality and fertility into the future, and those are usually based on a summary of the corresponding age-specific mortality and fertility in the past.  Hence there is an immediate need to represent full age schedules and their dynamics compactly.  This led to the widespread use of dimension-reduction or data compression techniques to reduce the dimensionality of the problem so that only a few parameters are necessary to represent age schedules and their dynamics.  \cite{ledermann1959dimensions} appear to have been the first to use principal components analysis (PCA) to summarize age-specific mortality and generate model life tables, and this approach was refined by many subsequent investigators, \citep[e.g.][]{bourgeois1962factor, bourgeois1990application, ledermann1969nouvelles, united1982model}. Following the early use of PCA to build model life tables, PCA and related methods like the singular value decomposition (SVD) \citep[e.g.][]{good1969some, stewart1993early, strang2009introduction} have been widely used and refined by forecasters to create time series models of mortality and fertility \citep[e.g.][]{bozik1987forecasting, lee1992modeling, lee1993modeling}.  \cite{bell1997comparing} provides a comprehensive summary of this line of development in various fields, dominated by actuarial science and applications in forecasting.

The `Lee-Carter' approach \citep{lee1992modeling, lee1993modeling} has been widely used in demography.  The model as presented in \cite{lee1992modeling} is %
\begin{align}
\mbox{ln}(\mbf{m}_{xt}) = \mbf{a}_x + \mbf{b}_xk_t + \epsilon_{xt} \ , 
\label{eq:leeCarter}
\end{align} %
where $x$ is age, $t$ is time, $\mbf{m}$ is a matrix of age, time-specific mortality rates, $\mbf{a}$ is the time-constant vector of mean (over columns of $\mbf{m}$) age-specific mortality rates through time, and $\mbf{b}$ is the time-constant first left singular vector from an SVD decomposition of the matrix of residuals generated by subtracting $\mbf{a}$ from each column of $\mbf{m}$. The expression can be rewritten as %
\begin{align}
\mbox{ln}(\mbf{m}_{xt}) = \bar{\mbf{a}}_x + \mbf{b}_xk_t \mid \bar{\mbf{a}}_x + \epsilon_{xt}  
\label{eq:leeCarterCond}
\end{align} %
to make clear that the $\mbf{b}_x k_t$ term models the column-wise residuals, or that fitting the model requires two separate steps: 
\begin{enumerate*}[label=\arabic*)]
\item calculate the residuals $r_{xt} = \mbox{ln}(\mbf{m}_{xt}) - \bar{\mbf{a}}_x$ and
\item \label{lc:last} extract the first left singular vector from the SVD of $\mbf{r}$ and calculate a $k_t$ value for each column of $\mbf{m}$ that minimizes the elements $\epsilon_{xt}$ for each column of $\mbf{m}$
\end{enumerate*}.

There are two conceptually separate elements to the Lee-Carter model, 
\begin{enumerate*}[label=\arabic*)]
\item a one-parameter (i.e. $k_t$) model of the full age-specific mortality or fertility schedule and
\item a time series model for that parameter
\end{enumerate*}.
The temporal sequence of values taken by $k_t$ is the focus of the time series model that is responsible for the temporal dynamics of the method, including the forecasts.  Development of the time series models is previewed in earlier work by the authors \citep{carter1986joint}.

The Lee-Carter model is a simplified version of a more complicated age-period-cohort mortality model conceived earlier by Wilmoth and elaborated over a number of years  \citep{wilmoth1987simple, wilmoth1989quand, wilmoth1990variation}\footnote{The core ideas underlying the Wilmoth model appear in his Ph.D. dissertation \citep{wilmoth1988Phd}, with further refinement in the following years, culminating in the English-language summary published in \textit{Sociological Methodology} in 1990 \citep{wilmoth1990variation}.}.  Wilmoth's model is designed to separate and identify age, period and cohort effects in an age $\times$ time matrix of mortality rates.  The basic structure is $\log(m_x) = [\mbox{mean model}] + [\mbox{residual model}]$ with the final form %
\begin{align}
f_{ij} = \underbrace{\alpha_i + \beta_j}_{\substack{\text{mean model}}} + \underbrace{\sum_{m=1}^\rho \phi_m \gamma_{im} \delta_{jm}}_{\substack{\text{1\textsuperscript{st} residual model}}} + \underbrace{\theta_k}_{\substack{\text{2\textsuperscript{nd}} \\ \text{residual} \\ \text{model}}} +\epsilon_{ij} \ , 
\label{eq:wilmoth}
\end{align} %
where $i$ is age, $j$ is period, $k = (j-i)$ indexes cohorts, $f$ is logged age-period-specific mortality $\log(m)$, $\alpha$ is an age effect, $\beta$ is a period effect, the sum $\sum_{m=1}^\rho \phi_{m} \gamma_{im} \delta_{jm}$ is over a set of $\rho$ rank-1 matrices from the SVD of the residuals remaining after the main effects are subtracted from $f$, and $\theta_k$ is a residual cohort effect remaining after subtracting both the main effects and the SVD approximation of the first residuals from $f$.  This form first appears in \cite{wilmoth1989quand}. 

The model is fit in three steps, effectively explaining ever more nuanced variation in a sequence of residuals.  As above, rewriting to the model with conditional terms as 
\begin{align}
f_{ij} = \alpha_i + \beta_j + \sum_{m=1}^\rho \phi_m \gamma_{im} \delta_{jm} \mid (\alpha_i + \beta_j) + \theta_k \mid \left( (\alpha_i + \beta_j) , \sum_{m=1}^\rho \phi_m \gamma_{im} \delta_{jm} \right) + \epsilon_{ij}  
\label{eq:wilmothCond}
\end{align}
may make this clear.  The three steps are: 
\begin{enumerate*}[label=\arabic*)]
\item calculate $\alpha_i$ and $\beta_j$ such that they minimize the first residuals $r_{ij} = f_{ij} - (\alpha_i + \beta_j)$, 
\item take the first $\rho$ terms from the SVD of the matrix of residuals $\mbf{r}$ and calculate the second residual $s_{ij} = r_{ij} - \sum_{m=1}^\rho \phi_m \gamma_{im} \delta_{jm}$ and 
\item calculate values for the elements of $\theta_k$ such that they minimize $s_{ij} - \theta_k = \epsilon_{ij}$
\end{enumerate*}.  
The SVD or `multiplicative' term  $\sum_{m=1}^\rho \phi_m \gamma_{im} \delta_{jm}$ takes shape over several publications \citep{wilmoth1987simple, wilmoth1989quand, wilmoth1990variation} to eventually be the standard SVD form that appears in the final model, with the first appearance of the SVD in \cite{wilmoth1989quand}.

A careful examination of Equations \ref{eq:leeCarter} and \ref{eq:wilmoth} reveals the relationship between the Wilmoth and Lee-Carter models. To move from Wilmoth to Lee-Carter:
\begin{enumerate*}[label=\arabic*)]
\item remove the main period effect $\beta_j$ and the cohort effect $\theta_k$ and
\item take only the first term in the SVD approximation of the first residual
\end{enumerate*}.
The SVD term then becomes $\phi_1 \gamma_{i1} \delta_{j1}$, or dropping the $m=1$ index, $\gamma_i (\phi \delta_j)$.  Replacing Wilmoth's $i$ and $j$ with Lee-Carter's $x$ and $t$ and letting $k = \phi \delta$ makes the equivalence transparent.  In their 1992 publication Lee and Carter acknowledge that their model has much in common with the Wilmoth model, but they do not correctly identify it as a simplified version of the Wilmoth model.  They go on to cite Wilmoth by way of explaining the SVD `solution' to calculating the elements of $\mbf{b}$, whereas again, this is just the simplest rank-1 form of the time-varying term in the model proposed by Wilmoth.  Consequently, the structure of the Lee-Carter model should be credited to Wilmoth, while Lee and Carter contribute the time series model for the time-varying elements of the Wilmoth model, namely the elements of the first right singular vector of the SVD of the mean-subtracted residuals, see below.

Motivated by the work of the UN Population Division that sometimes involves predicting full age schedules of mortality from child (and adult) mortality \citep{li2015wppLt}, \cite{wilmoth2012flexible} present another adaption of the original Wilmoth model, this time to generate model life tables as a function of $\qf$ or  $(\qf, \, \qff)$.  Adapting the nomenclature from log-linear models, this log-quadratic (Log-Quad) model has the form %
\begin{align}
\log(m_x) = a_x + b_xh + c_xh^2 + v_xk \ ,
\label{eq:logquad}
\end{align} %
where $x$ is age; $m$ is age-specific mortality; $a$,  $b$, and $c$ are constant age-specific coefficients for the quadratic mean model, $h$ is the input value of $\log(\qf)$, $v$ is an age-specific `correction factor', and $k$ is a coefficient for $v$.  Correction factor values $v_x$ are identified by calculating the SVD of the matrix of residuals that remain after the quadratic portion of the model is subtracted from life tables that are part Human Mortality Database  \citep{hmd2016} and using the resulting first left singular vector as a starting point\footnote{The first left singular vector of the HMD residuals are massaged slightly to ensure all elements of $v$ are positive and `smooth'.}.  Thus, the Log-Quad model has the now familiar mean/residual form of the original Wilmoth model and the structure of the residual model is a one-term version of the SVD form originally proposed by \cite{wilmoth1989quand}.  The Log-Quad's contribution is an innovative new mean model that takes advantage of the empirically observed curvilinear relationship between child mortality and mortality at other ages.  The Log-Quad model is elegant, simple, and parsimonious -- one ($\qf$) or two ($\qf$ and $k$)\footnote{If desired, $k$ is chosen so that the resulting mortality schedule matches an input value $\qff$.} parameters -- and it performs very well, accurately representing life tables with very low mortality and generally outperforming all other existing model life tables \citep{wilmoth2012flexible}.

Recently other investigators have worked on a variety of matrix-summary approaches to characterize the variability in mortality rates, but none of their work has been as widely used as the Wilmoth/Lee-Carter model.  Working independently, \cite{fosdick2012separable} develop an explicitly statistical `separable factor analysis' model to summarize mortality in the HMD, and at its core this is similar to the SVD term in Wilmoth's model.  

Also working independently, I developed a `component model' of mortality inspired by the use of matrix factorization methods and the fast Fourier transform in image compression.  The component model is a simple linear sum of independent, age-varying vectors (components) that when combined with appropriate weights can closely approximate age-specific mortality schedules.  This model has the simple basic form%
\begin{align}
\mbf{m} = \sum_{i=1}^\rho {w_i \mbf{u}_i} + \mbf{r}\ , 
\label{eq:compMod}
\end{align} %
where $\mbf{m}$ is a vector of age-specific mortality rates, $\mbf{u}_i$ are a set of vectors containing age-varying values identified by the SVD of a matrix of observed mortality rates, $w_i$ are weights, and $\mbf{r}$ is a vector of residuals.  This is similar to Ledermann's original use of factor analysis to build a system of model life tables based on factors resulting from a PCA decomposition of a matrix of age-specific mortality rates \citep{ledermann1959dimensions,ledermann1969nouvelles} and the PCA-based model underlying the UN model life tables \citep{united1982model} -- both of which have the mean/residual structure of the Wilmoth models because they use PCA operating on a centered data cloud. The component model has been used to summarize mortality data from the INDEPTH Network \citep{clarkPhD, indepthMLT2002, clark2009IndMltPaa}, similarly for the HMD \citep{clarkSharrow2011, clark2011ContMltPaa}, and more recently in work on small-area estimates of mortality \citep{alexander2016flexible}. This approach combines a simple linear model with PCA, SVD or similar methods to concentrate information along a few dimensions, see \citep{clark2015singular} for a detailed discussion.  

The component model is very similar to the SVD-inspired `1\textsuperscript{st} residual model' term in Wilmoth's Equation \ref{eq:wilmoth}.  However, neither Wilmoth nor subsequent investigators identify or develop the relationship between the SVD decomposition of a matrix of mortality rates and the column-wise, weighted-sum model  in Equation \ref{eq:compMod}.  A key conceptual difference between the two approaches is that Equation \ref{eq:compMod} does not have a `mean model', and consequently the factors identified by the SVD model everything, not just the residual as in all of the Wilmoth-inspired models.  The first component $\mbf{u}_1$ is effectively the mean age-specific mortality schedule and its weight reflects the overall level of mortality.  The remaining components $\mbf{u}_i$ for $i>1$ define deviations from the average age pattern, independent of level.  All of this follows directly from the properties of the SVD and a substantive interpretation of both the left and right singular vectors when applied to demographic age schedules \citep{clark2015singular}. Additionally, the weights are viewed as continuously varying parameters that can be the object or output of additional models - e.g. clustered using objective clustering methods to identify groups of similar age schedules, estimation using either traditional or Bayesian methods, or predicted from covariates that vary systematically with age schedules.

Finally, we recently applied the component model to HIV-related mortality in countries with large HIV epidemics \citep{sharrow2014modeling}.  In that article we demonstrate that the weights in Equation \ref{eq:compMod} vary systematically with HIV prevalence.  We took advantage of that fact to build a model that predicts three weights as a function of HIV prevalence and then predicts mortality age schedules from the predicted weights using Equation \ref{eq:compMod}.  The resulting `HIV-calibrated' component model uses the weights as a link between HIV prevalence and full age schedules of mortality.

\section{Data}

\subsection{Human Mortality Database Life Tables - HMD}

The Human Mortality Database (HMD) \citep{hmd2016} contains rigorously cleaned, checked and validated information on deaths and exposure from a number of developed countries.  The data are aggregated and presented in a wide variety of formats.  The objective of this analysis is to capture and characterize as much variability in age-specific mortality as possible, and consequently I chose to use the $1 \times 1$ HMD life tables for each sex.  Those provide all columns of a standard life table for single calendar years by single year of age from 0 $\rightarrow$ 110+.  Each country provides data for different historical periods, and some countries are subdivided into more specific subpopulations.  In the latter situation a `national population' life table is typically provided that aggregates across the subgroups.  Both the national and subgroup populations are included in this analysis to maximize the variability in age-specific mortality schedules in the overall dataset.  A few of the $1 \times 1$ life tables from the HMD contain problems: 
\begin{enumerate*}[label=\arabic*)]
\item the life tables for Belgium 1914-1918 for both sexes contain no data,
\item male life tables for Iceland (ISL) 1844, 1861, 1863, 1869, 1871, 1884, 1890, 1894 and New Zealand Mauri (NZL\_MA) 1958, 1979 display constant, generally implausibly low values for mortality at older ages, and likewise
\item female life tables for Iceland (ISL) 1852, 1864, 1882 and New Zealand Mauri (NZL\_MA) 1949, 1956, 1959, 1968 display similar implausible mortality at older ages
\end{enumerate*}.  All of those life tables are excluded.  Table \ref{tab:hmdData} contains an organized list of the life tables included in this analysis.  There are 4,486 life tables for each sex, 8,972 in total.  The HMD data used in this analysis are contained in the file at \url{http://www.mortality.org/hmd/zip/all_hmd/hmd_statistics.zip} downloaded on November 4, 2016.

\renewcommand{\arraystretch}{0.95}
\begin{table}[htp!]
\caption{\textbf{Life Tables.} 4,486 consistent $1 \times 1$ (single-year in both calendar and age) life tables downloaded from the Human Mortality Database on November 4, 2016.}
\begin{center}
\rowcolors{1}{}{lightgray}
\footnotesize
\begin{tabular}{ l l L{4.5cm}}

\toprule
Country (Code) & Subgroup (Code) & Years \\
\midrule
Australia (AUS) &  & 1921--2011 \\
Austria (AUT) &  & 1947--2014 \\
Belarus (BLR) &  & 1959--2014 \\
Belgium (BEL) &  & 1841--1913, 1919--2015 \\
Bulgaria (BGR) &  & 1947--2010 \\
Canada (CAN) &  & 1921--2011 \\
Chile (CHL) &  & 1992--2005 \\
Czech Republic (CZE) &  & 1950--2014 \\
Denmark (DNK) &  & 1835--2014 \\
Estonia (EST) &  & 1959--2013 \\
Finland (FIN) &  & 1878--2012 \\
France & Total population (FRATNP) & 1816--2014 \\
France & Civilian population (FRACNP) & 1816--2014 \\
Germany & Total population (DEUTNP) & 1990--2013 \\
Germany & East Germany (DEUTE) & 1956--2013 \\
Germany & West Germany (DEUTW) & 1956--2013 \\
Greece (GRC) &  & 1981--2013 \\
Hungary (HUN) &  & 1950--2014 \\
Iceland (ISL) &  & 1838--1843, 1845--1851, 1853--1860, 1862, 1865--1868, 1870, 1872--1881, 1883, 1885--1889, 1891--1893, 1895-- 2013 \\
Ireland (IRL) &  & 1950--2014 \\
Israel (ISR) &  & 1983--2014 \\
Italy (ITA) &  & 1872--2012 \\
Japan (JPN) &  & 1947--2014 \\
Latvia (LVA) &  & 1959--2013 \\
Lithuania (LTU) &  & 1959--2013 \\
Luxembourg (LUX) &  & 1960--2014 \\
Netherlands (NLD) &  & 1850--2012 \\
New Zealand & Total population (NZL\_NP) & 1948--2013 \\
New Zealand & Maori (NZL\_MA) & 1948, 1950--1955, 1957, 1960--1967, 1969--1978, 1980--2008 \\
New Zealand & Non-Maori (NZL\_NM) & 1901--2008 \\
Norway (NOR) &  & 1846--2014 \\
Poland (POL) &  & 1958--2014 \\
Portugal (PRT) &  & 1940--2012 \\
Russia (RUS) &  & 1959--2014 \\
Slovakia (SVK) &  & 1950--2014 \\
Slovenia (SVN) &  & 1983--2014 \\
Spain (ESP) &  & 1908--2014 \\
Sweden (SWE) &  & 1751--2014 \\
Switzerland (CHE) &  & 1876--2014 \\
Taiwan (TWN) &  & 1970--2014 \\
U.K. & United Kingdom Total Population (GBR\_NP) & 1922--2013 \\
U.K. & England \& Wales Total Population (GBRTENW) & 1841--2013 \\
U.K. & England \& Wales Civilian Population (GBRCENW) & 1841--2013 \\
U.K. & Scotland (GBR\_SCO) & 1855--2013 \\
U.K. & Northern Ireland (GBR\_NIR) & 1922--2013 \\
U.S.A. (USA) &  & 1933--2014 \\
Ukraine (UKR) &  & 1959--2013 \\

\bottomrule
\end{tabular}
\normalsize

\end{center}
\label{tab:hmdData}
\end{table}%
\renewcommand{\arraystretch}{1}

\subsection{Model Scales}

This analysis is conducted on life table probabilities of dying for those who survive to the beginning of each one-year age group.  Single year probabilities $\qox$ are taken directly from the HMD life tables, five-year probabilities $\qfx$ are calculated as $\qfx = 1 - \prod_{a=x}^{x+4}{(1-\qoa)}$, and $\qff$ is calculated as $\qff = 1 - \prod_{a=15}^{59}{(1-\qoa)}$.  `Child mortality' refers to $\qf$ and `adult mortality' refers to $\qff$.  

The natural scale of the models described below is the full real line, so life table probabilities of dying q are transformed using the \textit{logit} function $\logit(x) = \mbox{ln}\left(\frac{x}{1-x}\right)$ so that their transformed values occupy the full real line.   Outputs from the models are transformed back to the probability scale with range [0,1] using the \textit{expit} function $\expit(x) = \frac{\mbox{e}^x}{1+\mbox{e}^x}$, inverse of the \textit{logit}.

\section{Methods}

\subsection{Relevant Characteristics of the Singular Value Decomposition}

This section summarizes from \cite{clark2015singular}.  The SVD \citep{good1969some, stewart1993early,strang2009introduction} is a matrix factorization method that decomposes a matrix $\mbf{X}$ into three matrix factors with special properties: %
\begin{align}
\mbf{X}
&=
\mbf{USV}^\text{T} \ .
\label{eq:svd}
\end{align}
$\mbf{U}$ is a matrix of `left singular vectors' (LSVs) arranged in columns, $\mbf{V}$ is a matrix of `right singular vectors' (RSVs) arranged in columns, and $\mbf{S}$ is a diagonal matrix of `singular values' (SVs).  The LSVs and RSVs are independent and have unit length.  If one views the columns of $\mbf{X}$ as  a set of dimensions, then the rows of $\mbf{X}$ locate points defined by those dimensions -- the data cloud.  The RSVs define a new set of dimensions that line up with the axes of most variation in the data cloud.  The first RSV points from the origin to the data cloud, or if the cloud is around the origin, then it points along the line of maximum variation within the cloud.  The remaining RSVs are orthogonal to the first and each other and line up with successively less variable dimensions within the cloud.  The elements of the LSVs are values that correspond to each point along the new dimensions defined by the RSVs.  The SVs effectively stretch the new dimensions defined by the RSVs in accordance with the variation in the cloud along each RSV. 

The basic form of the SVD in Equation \ref{eq:svd} can be rearranged to yield two new useful expressions %
\begin{align*}
\mathbf{X} &= \sum_{i=1}^{\rho} s_{i} \mbf{u}_{i} \mbf{v}_{i}^\text{T} \Label{eq:sumRank-1}&
\mbox{and} & &
\mbf{x}_{\ell} &= \sum_{i=1}^{\rho} s_{i} v_{\ell i} \mbf{u}_{i} \ , \Label{eq:goldenEq}
\end{align*}%
where $\mbf{u}_i$ are LSVs, $\mbf{v}_i$ are RSVs, $s_i$ are SVs, $\rho$ is the rank of $\mbf{X}$, $\mbf{x}_\ell$ are columns of $\mbf{X}$, and $v_{\ell i}$ are the elements of RSV $\mbf{v}_i$, see App. \ref{app:svd}.  Equation \ref{eq:sumRank-1} says that $\mbf{X}$ can be written as a sum of rank-1 matrices, each created from one of the LSVs by applying weights in the form of the elements of the corresponding RSV.  Equivalently, Equation \ref{eq:goldenEq} says that each column $\mbf{x}_\ell$ of $\mbf{X}$ can be written as the weighted sum of the LSVs with the weight for each being the $\ell$\textsuperscript{th} element of the corresponding RSV\footnote{This is the expression used to model the first residual in Wilmoth's age/period/cohort model, Equation \ref{eq:wilmoth}.}.  The LSVs and SVs are constant, so the the weights are the `variables' in these expressions, and their values determine how much of each LSV is added to the mixture to represent the original data.  Finally because the LSVs are independent, OLS regression can be used to estimate models that relate $\mbf{x}_\ell$ to the LSVs. If the constant is constrained to be zero, then the coefficients are equal to $s_iv_{\ell i}$.

Because the RSVs define successively less variable dimensions in the data cloud, the first term in Equations \ref{eq:sumRank-1} and \ref{eq:goldenEq} contains the most information and subsequent terms contain less and less \citep{golub1987generalization}.  Including all $\rho$ terms replicates the original data matrix $\mbf{X}$ or any of its columns $\mbf{x}_\ell$ exactly, while including only the first few terms provides a good approximation.  Often in demographic applications only the first two to three terms are necessary for a close approximation, see \cite{clark2015singular}.

\subsection{SVD Component Model -- `SVD-Comp'}

Given an $A \times L$ matrix $\mbf{Q}$ of mortality schedules for each sex, calculate the $\mbox{SVD}(\mbf{Q}_z) = \mbf{U}_z \mbf{S}_z \mbf{V}_z^{\text{T}}$. Using the resulting factors as in Equation \ref{eq:goldenEq}, each mortality schedule $\mbf{q}_{z \ell}$ is approximated as the $c$-term sum%
\begin{align}
\mbf{q}_{z \ell} \approx \sum_{i=1}^{c} v_{z \ell i}  \cdot s_{z i} \mathbf{u}_{z i} \ ,
\label{eq:colRecon}
\end{align}%
where $z \in \{\mbox{female}, \mbox{male} \}$; $c \le \rho$, the rank of $\mbf{Q}_z$; and $\ell \in \{1 \dots L\}$ indexes mortality schedules \citep{golub1987generalization}. The LSVs $\mathbf{u}_{zi}$ and the SVs $s_{zi}$ are constant across all mortality schedules. Because $c \le \rho$, the sum on the right is an approximation of the mortality schedule, hence the `$\approx$'.  As is clear just below in Sec. \ref{sec:calibrate}, $c=4$ is sufficient to make the approximation almost perfect across the entire HMD\footnote{Viewed as a data compression technique, all 4,486 sex-specific mortality schedules in the HMD can be very closely approximated with just four age-varying components -- a nearly 99.9\%(!) reduction in the volume of data required to represent the HMD.}.  The elements that vary among mortality schedules are the RSVs $\mathbf{v}_{zi}$ whose elements $v_{z \cdot \, i}$ are the weights in the sum.  This is a continuously varying model like Mod-Logit \citep{murray2003} and Log-Quad \citep{wilmoth2012flexible} rather than a regional model like the Coale \& Demeny \citep{coale1966} and UN \citep{united1982model} model life tables.

When the $v_{z \ell i}$ are replaced by arbitrary values that can be related to covariates, as they are just below, this becomes a highly flexible modeling framework that can be used inductively like traditional model life tables to produce a mortality model that generates age schedules of mortality that are consistent with a collection of observed mortality schedules, or it can be used deductively to generate new age schedules based on a theoretical understanding of how a covariate should affect each component in the model.  In general, the age pattern of the scaled LSVs in the sum can be interpreted and manipulated theoretically, see Figure \ref{fig:svd} and the results in Sec. \ref{sec:results:SVDFacs}.

\subsection{Parameterization using $\qf$ and $(\qf, \, \qff)$} 

Equation \ref{eq:colRecon} describes a relationship between the elements of the RSVs and the age schedule of mortality.  Consequently, if a covariate is related to the age schedule of mortality, it will necessarily also have a relationship with the elements of the RSVs, particularly the first few RSVs corresponding to the SVD-defined dimensions that capture the majority of the variability in the data cloud formed by the HMD life tables.  It is possible to take advantage of this fact to define and estimate models that relate the elements of the RSVs to child mortality and adult mortality.  These take the form %
\begin{align*}
v_{z \ell i} &= f_{z i}(\qf_{\, z \ell}) \Label{eq:lsvC}& 
\mbox{and} & &
v_{z \ell i} &= f_{z i}(\qf_{\, z \ell}, \ \qff_{\, z \ell}) \ ,  \Label{eq:lsvCA}  
\end{align*} %
where, again, $z \in \{\mbox{female}, \mbox{male} \}$, $i \le \rho$ indexes the RSVs, and $\ell \in \{1 \dots L\}$ indexes both the elements of the RSVs and the values of child and adult mortality, one for each sex-specific mortality schedule. There is a separate model $f_{z i}$ for each sex-specific RSV, and these models can be used to produce predicted values for the weights in Equation \ref{eq:colRecon} using arbitrary values for  $\qf_{\, z}$ and $\qff_{\, z}$.

Following our earlier work \citep{sharrow2014modeling,indepthMLT2002}, the final model for an arbitrary set of age-specific mortality probabilities $\mbf{q}_z$ associated with given values for a set of weights $\widehat{w}_{z i} = f_{z i}(\qf_{\, z})$ or $\widehat{w}_{z i} = f_{z i}(\qf_{\, z}, \ \qff_{\, z})$ is%
\begin{align}
\widehat{\mbf{q}}_z = \sum_{i=1}^{c} \widehat{w}_{z i} \cdot s_{z i} \mathbf{u}_{z i}  \ .
\label{eq:colReconU5mr}
\end{align} %
Equation \ref{eq:colReconU5mr} relates either child mortality $\qf$ or both child and adult mortality $(\qf, \, \qff)$ to full age schedules of mortality according to the patterns of those relationship that exist in the original set of HMD life tables $\mbf{Q}$ using a very compact approximation.  

This is a fully general approach to predicting mortality, or any other, age schedules.  Equations \ref{eq:lsvC} and \ref{eq:lsvCA} can be replaced with models that summarize the relationships between any covariate and the RSVs and weights, and age can be aggregated into arbitrary age groups -- that simply requires recalculating the SVD on the age-aggregated data set.

\subsection{Calibrating SVD-Comp to the Relationship between $\qf$ and Mortality at Other Ages in the HMD \label{sec:calibrate}}

All computation is carried out using the \textsf{R} statistical programming environment \citep{RCore, R-url}.

\subsubsection{Calibration SVDs.} 

The life tables of the HMD are arranged into two $A \times L$ matrices $\mbf{Q}_z$ of single-year, age-specific life table probabilities of dying $\qox$, one for each sex. $A$ = number of age groups = 110, $L$ = number of life tables = 4,486, and $z \in \{ \mbox{female}, \, \mbox{male} \}$.  The SVD\footnote{SVDs calculated using the \textsf{svd} function in the \textsf{base} package of \textsf{R}.} of each $\mbf{Q}_z$ yields $\rho$ LSVs $\mbf{u}_{zi}$ and RSVs $\mbf{v}_{zi}$, and SVs  $\mbf{s}_z$. To ensure that all age groups have approximately the same influence when calculating the SVDs, each mortality schedule is offset from the origin\footnote{This ensures that the whole data cloud is separated from the origin by an amount that is substantially greater than the typical value of each logit-transformed mortality rate, and therefore each age group has roughly equivalent leverage in the optimization required to identify the first new dimension of the SVD.  The remaining dimensions are effectively identified on a centered data cloud.} by -10; the offset is added back to predicted mortality schedules.  Four of the new dimensions identified by each SVD are retained, i.e. $c=4$ in Equation \ref{eq:colReconU5mr}.  For females those account for 0.9983714, 9.216933e-04, 6.733335e-05, and 5.664095e-05 of the total sum of squares, respectively, or together 0.999417.  For males, 0.9986424, 8.043535e-04, 9.75966e-05, 4.971811e-05 and together 0.9995941.  

A final word about the SVs, the sum of the squares of the SVs is the total sum of squares in the original dataset (or cloud), so as either the number of points in the data cloud or the number of dimensions of the cloud increases, so will the total sum of squares and the values of the SVs, especially the first few.  Consequently, the scale of the SVs is dependent on the `size' of the dataset over which the SVD is calculated, and hence the scale of the components $s_{i} \mathbf{u}_{i}$ is also dependent on the size of the dataset.  In contrast the magnitude of the LSVs is constrained to be unity, but this means that the elements of the LSVs will be smaller as the number of elements increases, or as the number of points in the original dataset increases.  All this is to explain that the scale of the components is not fixed and depends on the size of the dataset over which the SVD is calculated.  Critically, this affects only the magnitude of the components, not their age patterns, and in practice none of this matters at all because the weights in Equation \ref{eq:colReconU5mr} can incorporate a factor that accounts for scale.

\subsubsection{Models for Predicting Weights. \label{sec:modsPredWeights}} 

Based on Equations \ref{eq:lsvC} and \ref{eq:lsvCA}, regression models are defined that relate the RSVs $\mbf{v}_{zi}$ to $\qf_{\, z}$ and $\qff_{\, z}$.  Scatterplots of the elements of the RSVs versus $\logit(\qf)$ in Figures \ref{fig:rsvF} and \ref{fig:rsvM} make it clear that the relationships are not linear or simple.  With no theory to guide the choice of predictors, I tried all combinations of simple transformations of $\logit(\qf)$ and $\logit(\qff)$ and their interactions.  The resulting models explain almost all the variance in the elements of $\mbf{v}_1$ ($\mbox{R}^2 \approx 98\%$ for both sexes), the vast majority of the variance in the elements of $\mbf{v}_2$ ($\mbox{R}^2 \approx 87\%$ for both sexes), and between one third and one half of the variance in the elements of $\mbf{v}_3$ an d $\mbf{v}_4$.  Additionally, I tried to avoid overfitting or creating odd boundary effects in the predicted values that would have made out-of-sample predictions immediately implausible.  These models behave sensibly up to the edges of the sample. The final models are %
\begin{align}
v_{z \ell i} = c_{zi} &+  \beta_{z1i} \cdot \qf_{\, z\ell} + \beta_{z2i} \cdot \logit(\qf)_{\, z\ell} + \beta_{z3i} \cdot \logit(\qf)^2_{\, z\ell} + \beta_{z4i} \cdot \logit(\qf)^3_{\, z\ell} \nonumber \\  
 &+ \beta_{z5i} \cdot \qff_{\, z\ell} + \beta_{z6i} \cdot \logit(\qff)^2_{\, z\ell} +  \beta_{z7i} \cdot \logit(\qff)^3_{\, z\ell} \nonumber \\ 
 &+ \beta_{z8i} \cdot [\logit(\qf)_{\, z\ell} \times \logit(\qff)_{\, z\ell}] + \epsilon_{z\ell i} \ , \label{eqn:vsByMx} 
\end{align}%
where $i \in \{1:4\}$ indexes the SVD dimensions and $\ell$ indexes mortality schedules and elements of $\mbf{v}_{zi}$.  OLS regression is used to estimate coefficients for the eight regression models defined in Equation \ref{eqn:vsByMx}, and the estimated values are contained in App. A Tables \ref{tab:appA:femaleRSVMods} and \ref{tab:appA:maleRSVMods}.  Using arbitrary values for both $\qf$ and $\qff$ as inputs, these models are used to predict values for the weights in Equation \ref{eq:colReconU5mr}.

\subsubsection{Models for Adult Mortality. \label{sec:modsAdultMx}} 

To accommodate a one-parameter model that uses only $\qf$ as an input, a regression model is defined that relates adult mortality $\logit(\qff)_{z}$ to child mortality $\qf_{\, z}$.  The scatterplot of $\logit(\qff)$ versus $\logit(\qf)$ in Figure \ref{fig:adultChild} reveals a slightly complicated relationship that is neither linear nor systematically curvilinear.  Again without theory as a guide, I tried a variety of models including various simple transformations of $\qf$.  The resulting models explain almost all the variance in $\logit(\qff)$ ($\mbox{R}^2 = 93\%$ for females and $79\%$ for males).  The final models are%
\begin{align}
\logit(\qff)_{\, z\ell} = c_z &+ \beta_{z1} \cdot {\qf}_{\, z\ell} + \beta_{z2} \cdot \logit(\qf)_{\, z\ell} \nonumber \\
&+ \beta_{z3} \cdot \logit(\qf)^2_{\, z\ell} + \beta_{z4} \cdot \logit(\qf)^3_{\, z\ell} + \epsilon_{z\ell} \ . \label{eqn:amCm}
\end{align}%
OLS regression is used to estimate coefficients for the two regression models defined by Equation \ref{eqn:amCm}, and the estimated coefficients are contained in App. A Table \ref{tab:appA:adultMxMod}.  This model is used to predict values for $\qff$ when only $\qf$ is supplied as an input.  Then both the input value for $\qf$ and the predicted value for $\qff$ are used in Equation \ref{eqn:vsByMx} to predict the weights in Equation \ref{eq:colReconU5mr}.   

\subsubsection{Models for Mortality in the First Year of Life. \label{sec:modsInfMx}} 

Mortality falls very rapidly in the first few years of life.  Using the child mortality rate $\qf$, a five-year summary of mortality between ages 0 and 5, as a predictor of single-year mortality within that same five-year age group is relatively uninformative.  Experimentation reveals that $\qf$ predicts $\qoo$ through $\qof$ well and $\qoz$ slightly less well.  The prediction of $\qoz$ can be improved by modeling the relationship between $\logit(\qoz)$ and $\logit(\qf)$ separately as 
\begin{align}
\logit(\qoz)_{z\ell} = c_z + \beta_{z1} \cdot \logit(\qf)_{\, z\ell} + \beta_{z2} \cdot \logit(\qf)^2_{\, z\ell} + \epsilon_{z\ell} \ . \label{eqn:q0Cm}
\end{align}
OLS regression is used to estimate the coefficients of this model, displayed in App A. Table \ref{tab:appA:infantMxMod}.  The model explains essentially all the variance in $\logit(\qoz)$ ($\mbox{R}^2 > 99\%$ for both sexes) and is used to predict values for $\qoz$ directly from the input value of $\qf$.

\subsection{Using the Model}

The full model is used in the following way:

\begin{enumerate}

\item \label{en:use1} Identify input values for $\qf$ and optionally $\qff$ and transform them to the logit scale.  If $\qff$ is not available, predict $\logit(\qff)$ using the input value for $\qf$ and the regression coefficients corresponding to Equation  \ref{eqn:amCm}.

\item \label{en:use2} Use the input values for $\logit(\qf)$ and $\logit(\qff)$ obtained in step \ref{en:use1} and the regression coefficients corresponding to Equation \ref{eqn:vsByMx} to predict values for the weights $\widehat{w}_{zi}$ defined Equation \ref{eq:colReconU5mr}.  

\item Insert the weights predicted in step \ref{en:use2} into Equation \ref{eq:colReconU5mr}  to calculate a predicted age schedule of mortality probabilities $\widehat{\mbf{q}}$ on the logit scale.  

\item If desired, improve the prediction of $\logit(\qoz)$ using the regression coefficients  corresponding to Equation \ref{eqn:q0Cm} to directly predict $\logit(\qoz)$ from the input value of $\qf$ from step \ref{en:use1}.  Replace the first element of $\widehat{\mbf{q}}$ with this predicted value for $\logit(\qoz)$.

\item Add 10 to each element of $\widehat{\mbf{q}}$ to account for the offset used when calculating the SVDs of the HMD mortality schedules.

\item Take the expit of $\widehat{\mbf{q}}$ to yield single-year age-specific probabilities of dying on the probability scale.

\end{enumerate}

\subsection{Model Validation}

The general sensitivity of the model to exactly which mortality schedules are used for calibration is assessed using a cross validation approach.  Twenty-five random samples of 50\% of the HMD mortality schedules are drawn, the model is calibrated with each using the calibration process described just above in Sec. \ref{sec:calibrate}, and all of the HMD mortality schedules are predicted.  For each of the 25 models, prediction errors are calculated for all mortality schedule as the difference $\mbf{q}_\ell - \widehat{\mbf{q}}_\ell$. The error distributions of the in-sample and out-of-sample mortality schedules are summarized and compared.

In order to investigate how sensitive the overall modeling approach is to the number of mortality schedules used to calibrate the model, another cross validation exercise is conducted with varying sample sizes.  For each sample fractions from 10\% to 90\% in 20\% increments, 50 random samples are drawn from the HMD life tables .  As above, the model is calibrated using each sample and all of the HMD mortality schedules are predicted, errors calculated, and error distributions for in- and out-of-sample mortality schedules are summarized and compared.

\subsection{Comparing Performance of SVD-Comp and the Log-Quad Model}

The Log-Quad model \citep{wilmoth2012flexible} is the state of the art mortality model relating child and adult mortality to full age schedules of mortality.  I compare prediction errors produced by both the Log-Quad and SVD-Comp models.  For the Log-Quad model I use \textsf{R} code provided by \cite{wilmoth2012flexible} to produce predicted $\qfx$ values for each of the HMD mortality schedules using either  $\qf$ or both  $\qf$ and  $\qff$ as inputs.  The Log-Quad model predicts mortality in five-year age groups.  To accommodate that using the one-year age groups ($\qox$) predicted by the SVD-Comp model, I use standard life table methods to transform predicted single-year to five-year $\qnx$ values.  I summarize the distribution of errors $\mbf{q}_\ell - \widehat{\mbf{q}}_\ell$ produced by both models in various ways.  Comparisons are made only for predictions using the same inputs for both models, either $\qf$ alone or both $(\qf, \, \qff)$. 

I also summarize the overall error produced by each model across all of the mortality schedules in the HMD.  This is done by taking the absolute value of each year-sex-age-specific error and then summing the resulting absolute errors across all ages and years for each sex.  This produces a single number -- the total absolute error -- that indicates the overall difference between the predicted and actual values for all years and ages. 

\section{Results \label{sec:results}}

\subsection{Data and Fits}

\begin{figure}[htbp]
   \centering
   \includegraphics[width=\linewidth]{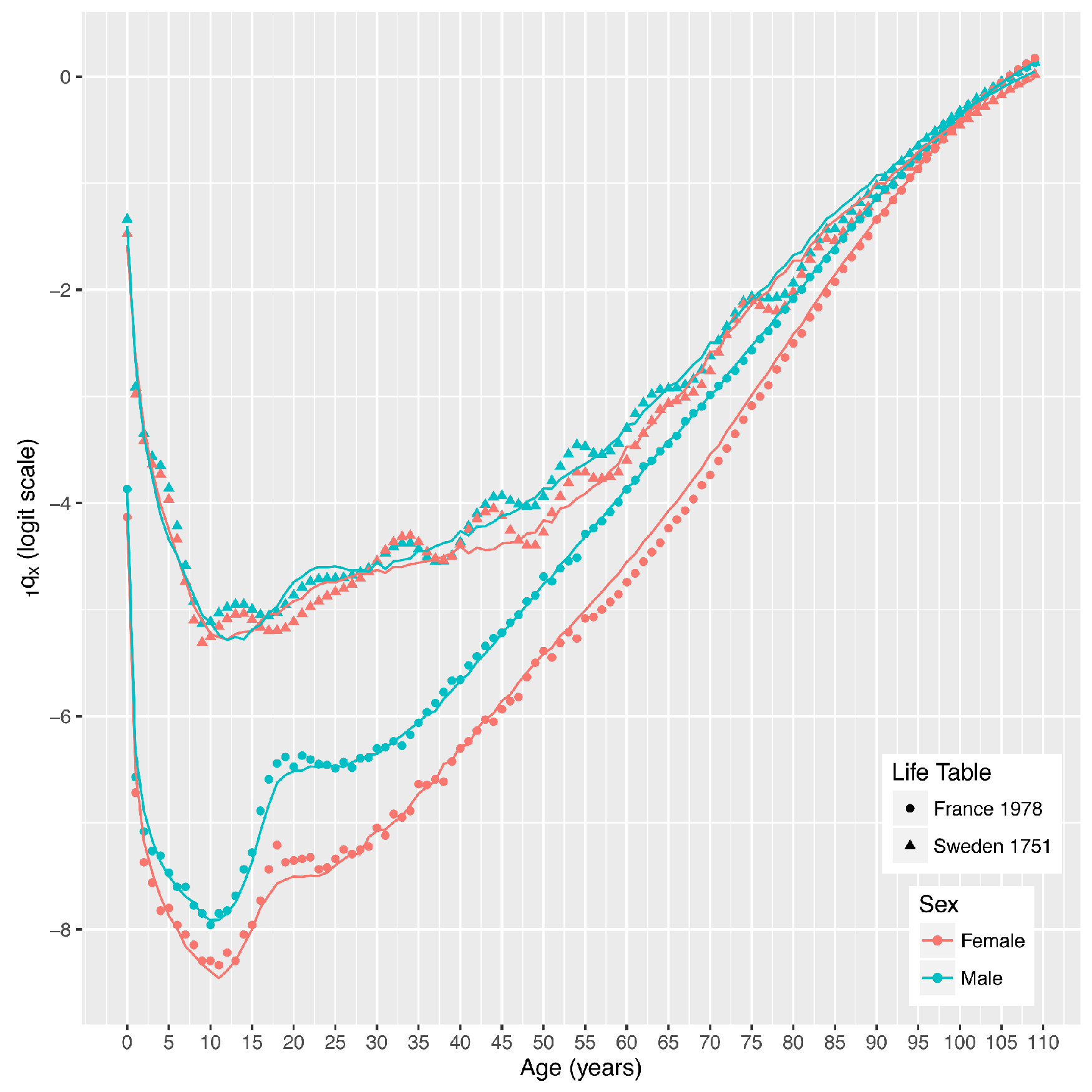} 
   \captionsetup{format=plain,font=normalsize,margin=0cm,justification=justified}
   \caption{\textbf{Example Data \& Predictions.}  $\qox$ for very high mortality early in Sweden's time series and low mortality for a more recent year in France.  Predicted values produced using $\qf$ alone as an input.  Data as symbols and predicted values as lines.}
   \label{fig:data}
\end{figure}

To provide a sense of the mortality data contained in the HMD and the fits produced by the SVD-Comp model, Figure \ref{fig:data} displays $\qox$ on the logit scale for Sweden in 1751 and France in 1978, with both data and predicted values produced by SVD-Comp using $\qf$ alone as an input.

\subsection{Factors of the SVD \label{sec:results:SVDFacs}}

\begin{figure}[htbp]
   \centering
   \includegraphics[width=\linewidth]{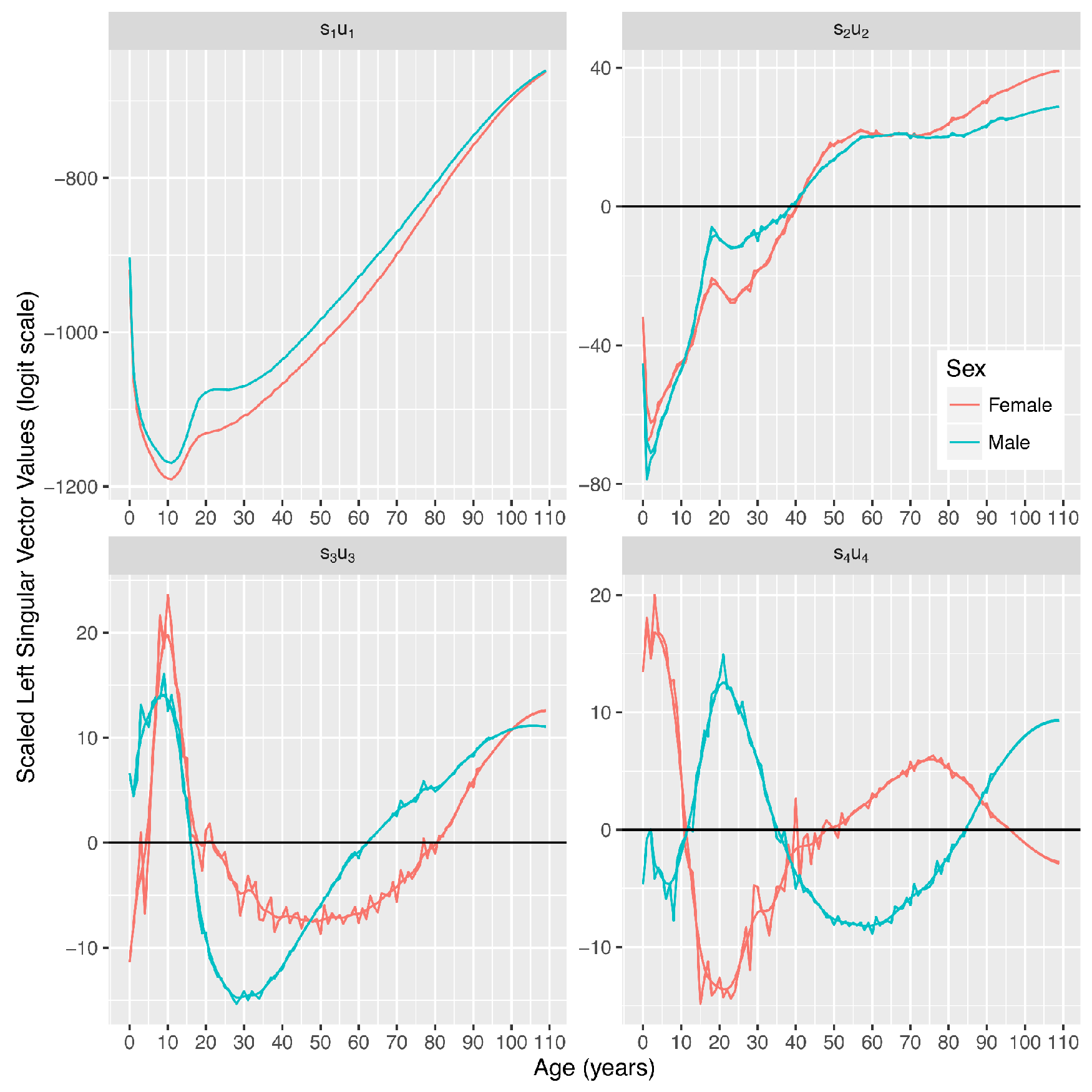} 
   \captionsetup{format=plain,font=normalsize,margin=0cm,justification=justified}
   \caption{\textbf{Scaled Left Singular Vectors.} First four LSVs scaled by their corresponding SVs from the SVD of the 4,486 mortality schedules in the HMD.}
   \label{fig:svd}
\end{figure}

Figure \ref{fig:svd} presents the sex-specific LSVs from the SVD of the full set of HMD mortality schedules scaled by their corresponding singular values, $s_{i} \mathbf{u}_{i}$ (ignoring the index for sex $z$).  All elements of $s_1\mbf{u}_1$ are negative so that $s_1\mbf{u}_1$ captures the underlying `average' shape of the mortality profile with age.  Weights applied to $s_1\mbf{u}_1$ move this underlying mortality profile up and down and hence control the overall level of mortality.  The remaining $s_i\mbf{u}_i$ all cross zero and therefore represent age-specific deviations from the overall underlying pattern.  These scaled left singular vectors are the components used in the weighted sum in Equation \ref{eq:colReconU5mr}.  Figure \ref{fig:svd} also displays smoothed\footnote{Kernel smoother with guassian kernal and bandwidth = $i+1$ for ages older than $i$.} versions of the scaled LSVs.  One can use the smoothed versions to make the predicted mortality schedules smoother.

\subsection{Calibration Relationships}

\begin{figure}[htbp]
   \centering
   \includegraphics[width=\linewidth]{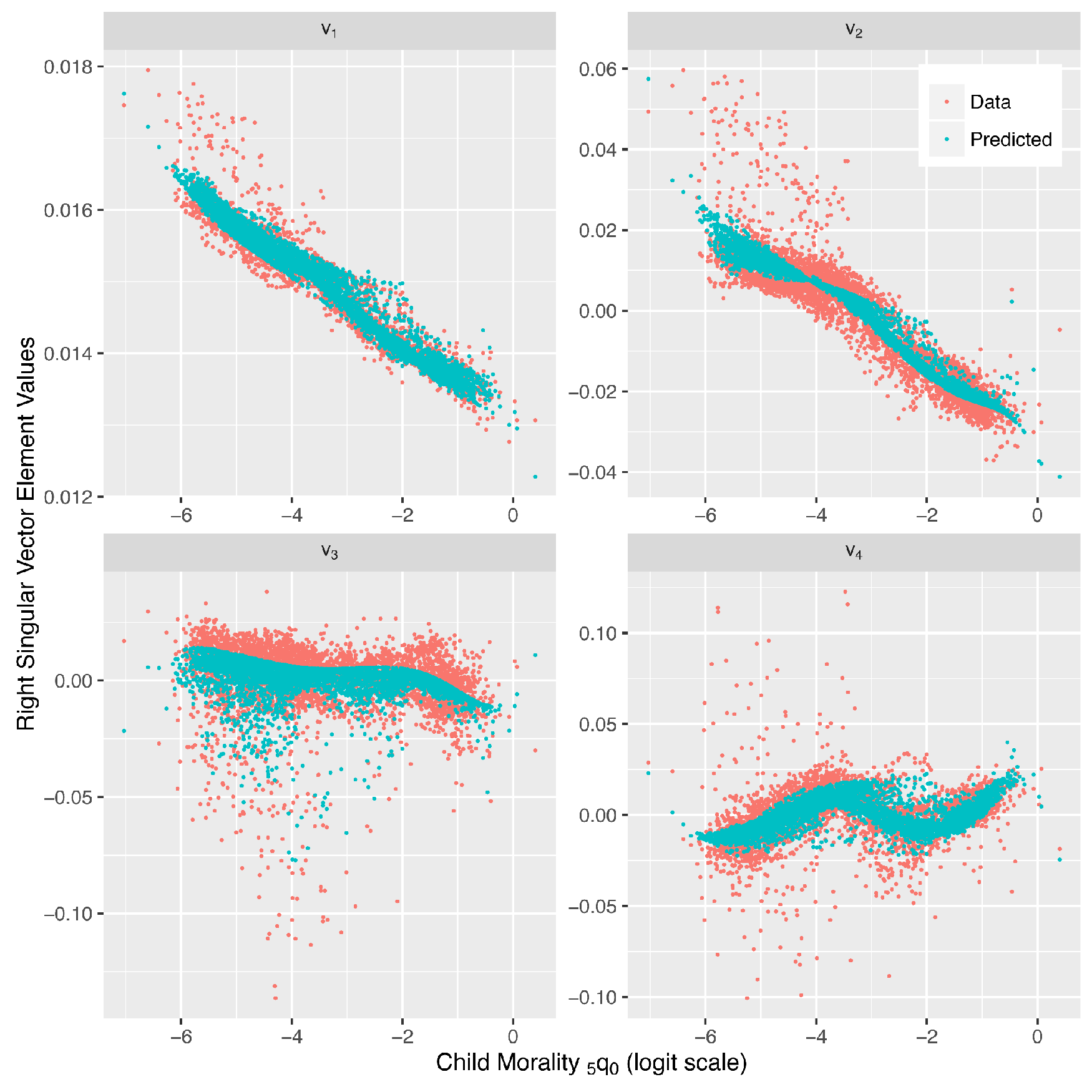} 
   \captionsetup{format=plain,font=normalsize,margin=0cm,justification=justified}
   \caption{\textbf{Right Singular Vector Element Values for Females.} Values and predictions from model in Equation \ref{eqn:vsByMx} on the logit scale by $\logit(\qf)$.  The predicted values are based on both $\qf$ and $\qff$ which explains why they appear as a cloud rather than a curve.}
   \label{fig:rsvF}
\end{figure}

\begin{figure}[htbp]
   \centering
   \includegraphics[width=\linewidth]{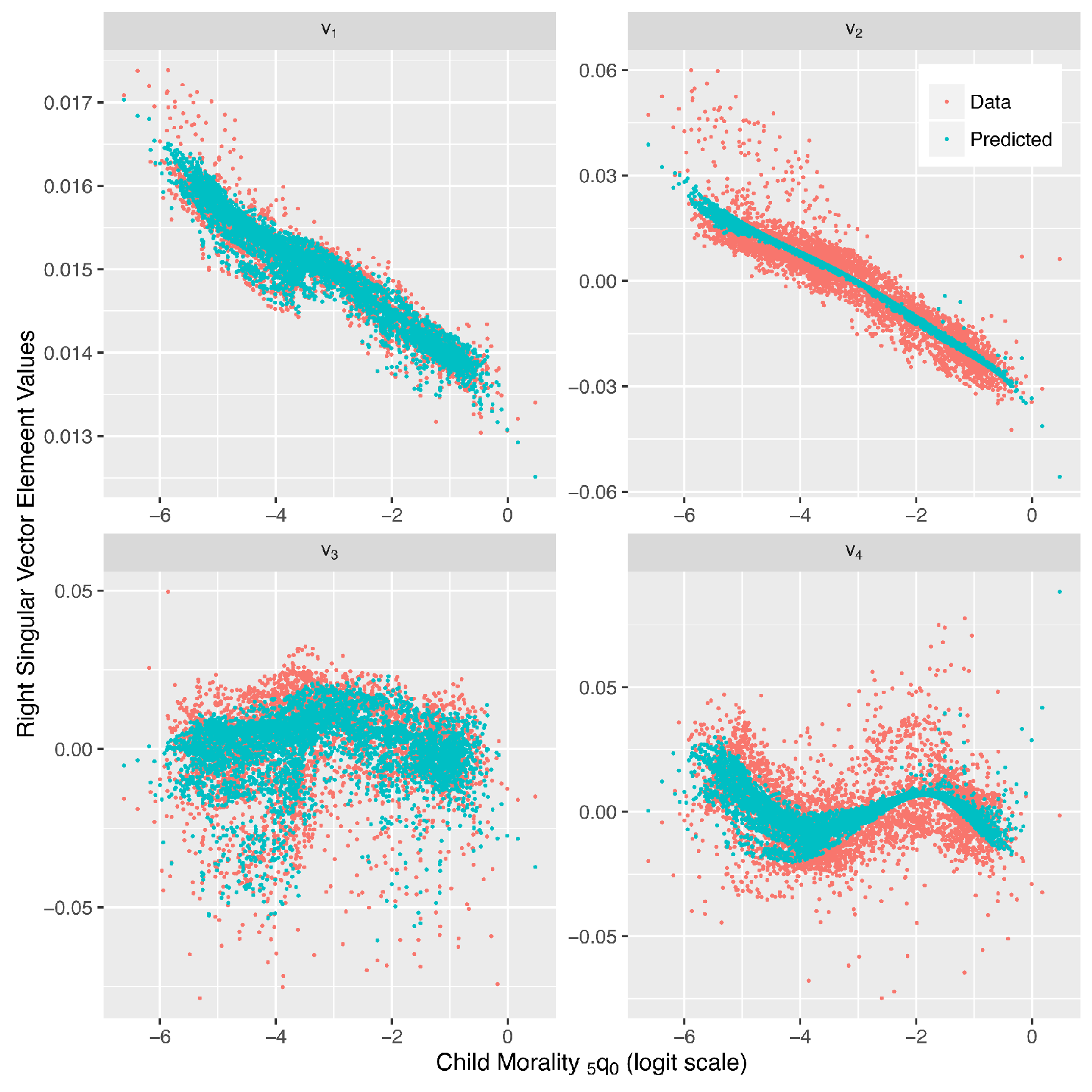} 
   \captionsetup{format=plain,font=normalsize,margin=0cm,justification=justified}
   \caption{\textbf{Right Singular Vector Element Values for Males.} Values and predictions from model in Equation \ref{eqn:vsByMx} on the logit scale by $\logit(\qf)$.  The predicted values are based on both $\qf$ and $\qff$ which explains why they appear as a cloud rather than a curve.}
   \label{fig:rsvM}
\end{figure}

\begin{figure}[htbp]
   \centering
   \includegraphics[width=\linewidth]{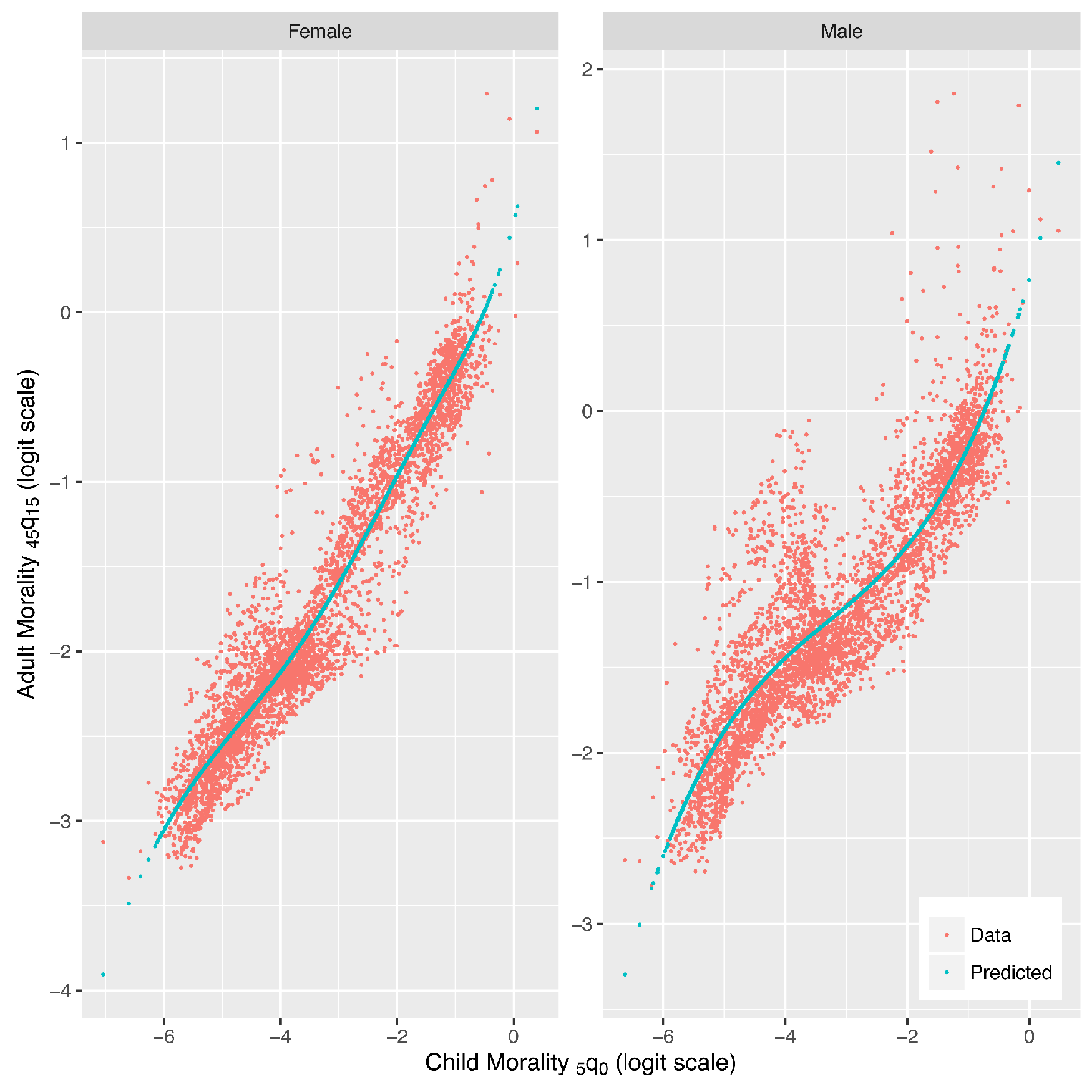} 
   \captionsetup{format=plain,font=normalsize,margin=0cm,justification=justified}
   \caption{\textbf{Adult vs. Child Mortality.} Values and predictions from model in Equation \ref{eqn:amCm} on the logit scale by $\logit(\qf)$.}
   \label{fig:adultChild}
\end{figure}

\begin{figure}[htbp]
   \centering
   \includegraphics[width=\linewidth]{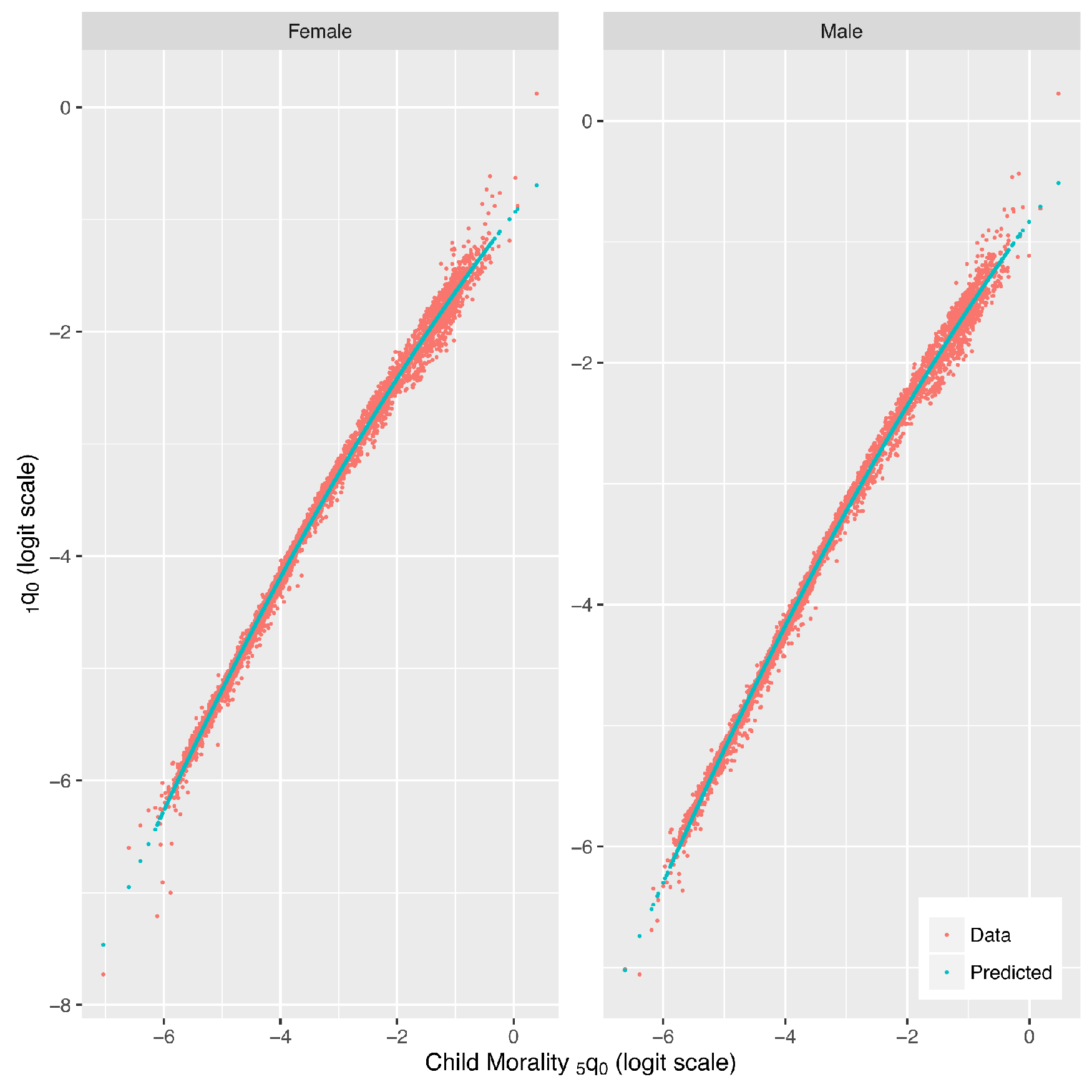} 
   \captionsetup{format=plain,font=normalsize,margin=0cm,justification=justified}
   \caption{\textbf{Age 0 Probability of Dying $\qoz$ vs. Child Mortality.} Values and predictions from model in Equation \ref{eqn:q0Cm} on the logit scale by $\logit(\qf)$.}
   \label{fig:q0Child}
\end{figure}

Figures \ref{fig:rsvF} through \ref{fig:q0Child} display the data and predicted values from the models in Equations \ref{eqn:vsByMx}, \ref{eqn:amCm}, and \ref{eqn:q0Cm}, and the corresponding estimated coefficients based on the whole HMD and used to calculate the predicted values in the figures are contained in Tables \ref{tab:appA:femaleRSVMods}, \ref{tab:appA:maleRSVMods}, \ref{tab:appA:adultMxMod}, and \ref{tab:appA:infantMxMod} in Appendix \ref{app:regs}.  Figures \ref{fig:rsvF} and \ref{fig:rsvM} contain scatterplots of the RSV element values versus $\logit(\qf)$.  The figures display both data and values predicted from Equation \ref{eqn:vsByMx} using $\logit(\qf)$ and $\logit(\qff)$ predicted from the model in Equation \ref{eqn:amCm} as inputs.  There are clear, quasilinear relationships between the elements of $\mbf{v}$s and $\logit(\qf)$. Figure \ref{fig:adultChild} displays $\logit(\qff)$ versus $\logit(\qf)$, along with the predicted values from Equation \ref{eqn:amCm}.  Finally, Figure \ref{fig:q0Child} displays $\qoz$ versus $\logit(\qf)$, along with predicted values from Equation \ref{eqn:q0Cm}.

\subsection{Cross Validation Prediction Errors}

\begin{figure}[htbp]
   \centering
   \includegraphics[width=\linewidth]{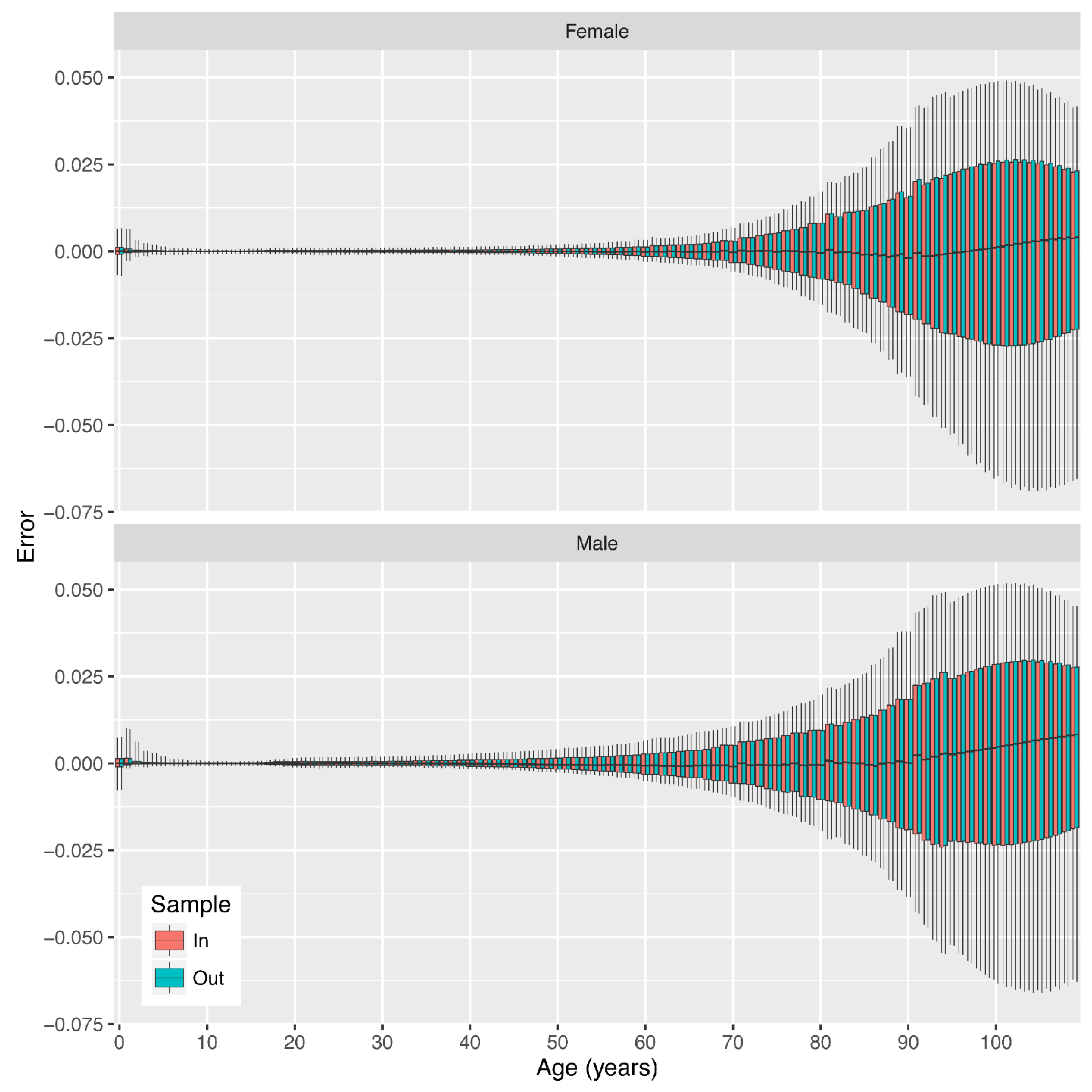} 
   \captionsetup{format=plain,font=normalsize,margin=0cm,justification=justified}
   \caption{\textbf{SVD-Comp Prediction Errors.}  Single-year age group prediction errors for in- and out-of-sample mortality schedules.  25 50\% samples.  Errors summarized over all in- and out-of-sample mortality schedules for the 25 samples, each box summarizes 56,075 errors. Whiskers extend to 10\% and 90\% quantiles.}
   \label{fig:ageSpecErr}
\end{figure}

Figure \ref{fig:ageSpecErr} displays sex-age-specific boxplots of the error distribution for one-year age groups from the first cross validation using 25 samples of 50\% of the HMD to calibrate the SVD-Comp model.  The errors are generally very small and centered around zero through roughly age 60.  At older ages the size of the errors increases, and the median drifts slightly away from zero in a positive direction, especially at ages older than 90.  However, the median error is never more than 0.025, and as displayed in Figure \ref{fig:svdVsLqC}, they are significantly smaller than the median errors produced by the Log-Quad model at the same ages.  The error distributions of the in- and out-of-sample predictions are indistinguishable at all ages indicating that the SVD-Comp model is not sensitive to exactly which mortality schedules are used for calibration when half of them are used.

\subsection{Varying Sample Size Cross Validation Prediction Errors}

\begin{figure}[htbp]
   \centering
   \includegraphics[width=\linewidth]{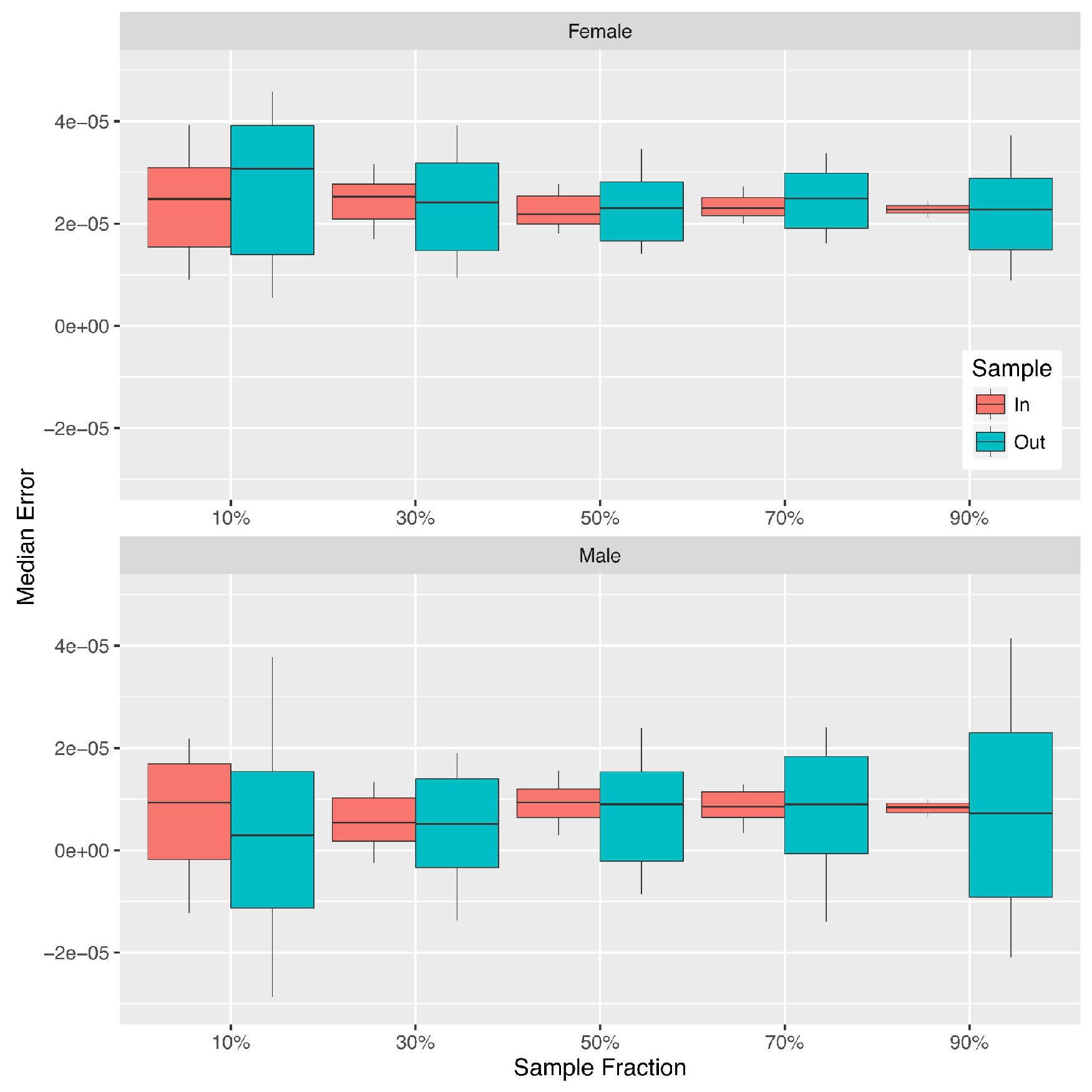} 
   \captionsetup{format=plain,font=normalsize,margin=0cm,justification=justified}
   \caption{\textbf{Median Prediction Error by Sample Fraction.}  50 samples for each sample fraction.  For each sample, median calculated across all ages and all mortality schedules in each sample category (in/out), boxplots summarize 50 values for the median, one for each sample. Whiskers extend to 10\% and 90\% quantiles.}
   \label{fig:medSampErr}
\end{figure}

\begin{figure}[htbp]
   \centering
   \includegraphics[width=\linewidth]{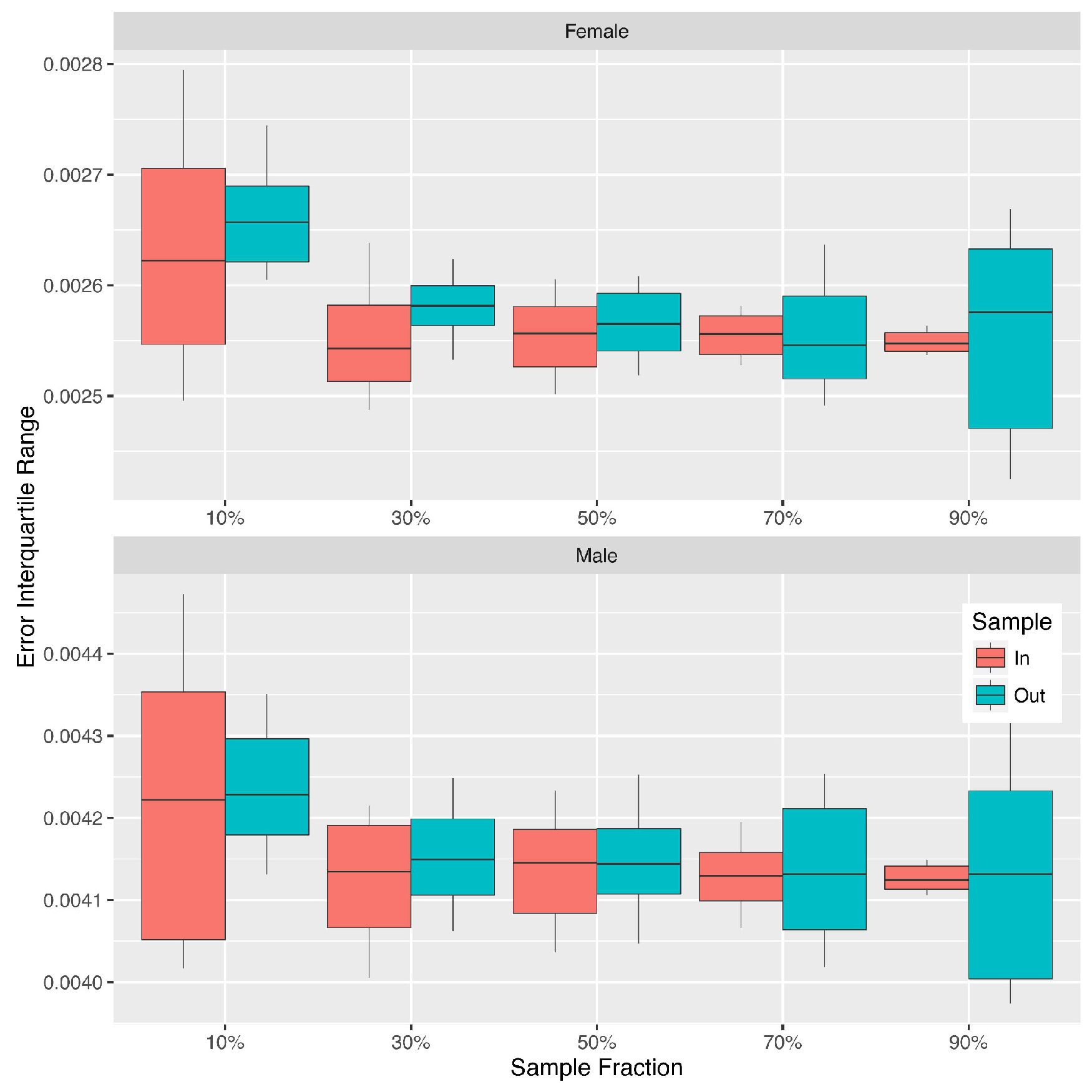} 
   \captionsetup{format=plain,font=normalsize,margin=0cm,justification=justified}
   \caption{\textbf{Interquartile Range of Prediction Error by Sample Fraction.}  50 samples for each sample fraction.  For each sample, the interquartile range is calculated across all ages and all mortality schedules in each sample category (in/out), boxplots summarize 50 values for the interquartile range, one for each sample.  Whiskers extend to 10\% and 90\% quantiles.}
   \label{fig:iqrSampErr}
\end{figure}

Figures \ref{fig:medSampErr} and \ref{fig:iqrSampErr} contain the second set of cross validation results investigating the effect of varying the number of mortality schedules used to calibrate the SVD-Comp model.  Both figures summarize the overall prediction error distributions (all ages and years combined) for the SVD-Comp model by sample status, in- versus out-of-sample mortality schedules.  The sample fraction varies from 10\% to 90\% in increments of 20\%.   Figure \ref{fig:medSampErr} displays boxplots of the median overall error.  The median of median overall errors is very similar comparing in- and out-of-sample mortality schedules for both sexes across all sample fractions.  There is a slight positive bias in all cases resulting from the positive bias in errors at older ages, see Figure \ref{fig:ageSpecErr}.  A similar situation exists for the distributions of the interquartile range of overall errors, Figure \ref{fig:iqrSampErr}.  The only systematic change in these distributions by sample fraction is that the interquartile range of the indicators calculated from the sample decreases as the sample fraction increases, as expected.  Inversely, there is a weak trend toward increases in the interquartile range calculated in the out-of-sample group as the sample fraction increases, also as expected.  In general the SVD-Comp model appears to be remarkably robust as the number of mortality schedules used for calibration decreases.  Performance is satisfactory all the way down to the 10\% sample and good all all the way down to 30\%.

\subsection{Comparison between SVD-Comp and Log-Quad Prediction Errors}

\begin{figure}[htbp]
   \centering
   \includegraphics[width=\linewidth]{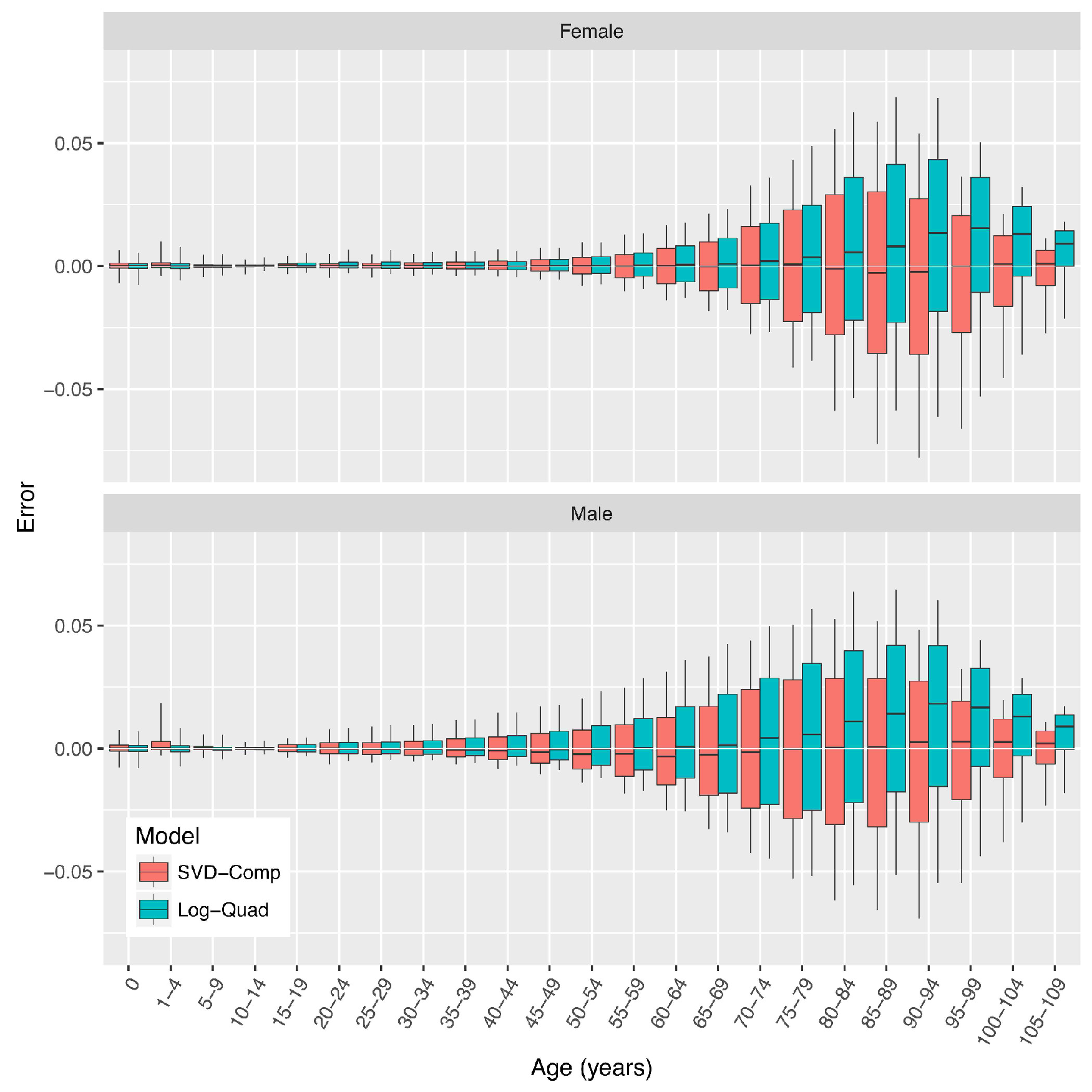} 
   \captionsetup{format=plain,font=normalsize,margin=0cm,justification=justified}
   \caption{\textbf{SVD-Comp and Log-Quad Prediction Errors.}  Five-year age group prediction errors for SVD-Comp and Log-Quad models using only child mortality $\qf$ as input.  Each box summarizes 4,486 errors. Whiskers extend to 10\% and 90\% quantiles.}
   \label{fig:svdVsLqC}
\end{figure}

Figure \ref{fig:svdVsLqC} displays sex-age-specific boxplots of the distribution of prediction errors for both the SVD-Comp and Log-Quad models.  The median error by sex and age is close to zero for both models through roughly age 70.  At ages older than 70 the median error for the Log-Quad model is systematically significantly larger than zero, while for the SVD-Comp model the median error stays at zero.  The sex-age-specific interquartile ranges are similar for both models, very small through roughly age 40, growing slowly between 40 and roughly 85 and then shrinking again through 110. In general at ages older than 45 the error distribution for the Log-Quad model is biased in a positive direction, while for the SVD-Comp model the error distribution is centered around zero at all ages.

Table \ref{tab:absErrs} displays the total absolute errors for the SVD-Comp and Log-Quad models for predictions based on either $\qf$  alone or both $(\qf, \, \qff)$.  The table also presents differences between the total absolute errors for the two models in both additive (Log-Quad - SVD-Comp) and proportional form ([Log-Quad - SVD-Comp]/SVD-Comp).  In all cases the SVD-Comp model predictions are globally closer to the truth.

\begin{table}[H]
\captionsetup{format=plain,font=normalsize,margin=2.9cm,justification=justified}
\caption{\textbf{Summary of Prediction Errors for SVD-Comp and Log-Quad.}  Total absolute error and comparisons of total absolute error.  Both models trained on all HMD life tables.}
\begin{center}
\begin{tabular}{llrrr}
\toprule
  &   & \multicolumn{3}{c}{Total Absolute Error Predicted by} \\
\cmidrule{3-5}
 &   & \multicolumn{1}{c}{C1} & \multicolumn{1}{c}{C2} & \multicolumn{1}{c}{C3} \\
 & Model / Summary  & \multicolumn{1}{c}{$\qf$} & \multicolumn{1}{c}{$(\qf, \, \qff)$} & \multicolumn{1}{c}{C2-C1} \\
 \midrule
\multicolumn{5}{l}{\textit{Female}} \\
 \midrule
R1 & SVD-Comp & 1,386 & 1,244 & -142 \\
R2 & Log-Quad & 1,439 & 1,339 & -100 \\
R3 & R2-R1 & 53 & 95 & 42 \\
R4 & R3/R1 (\%) & 3.8\% & 7.6\% & 3.8\% \\
 \midrule
\multicolumn{5}{l}{\textit{Male}} \\
 \midrule
R5 & SVD-Comp & 1,595 & 1,308 & -287 \\
R6 & Log-Quad & 1,691 & 1,400 & -291 \\
R7 & R6-R5 & 96 & 92 & -4 \\
R8 & R7/R5 (\%) & 6.0\% & 7.0\% & 1.0\% \\
\bottomrule
\end{tabular}
\end{center}
\label{tab:absErrs}
\end{table}

\begin{figure}[htbp]
   \centering
   \includegraphics[width=\linewidth]{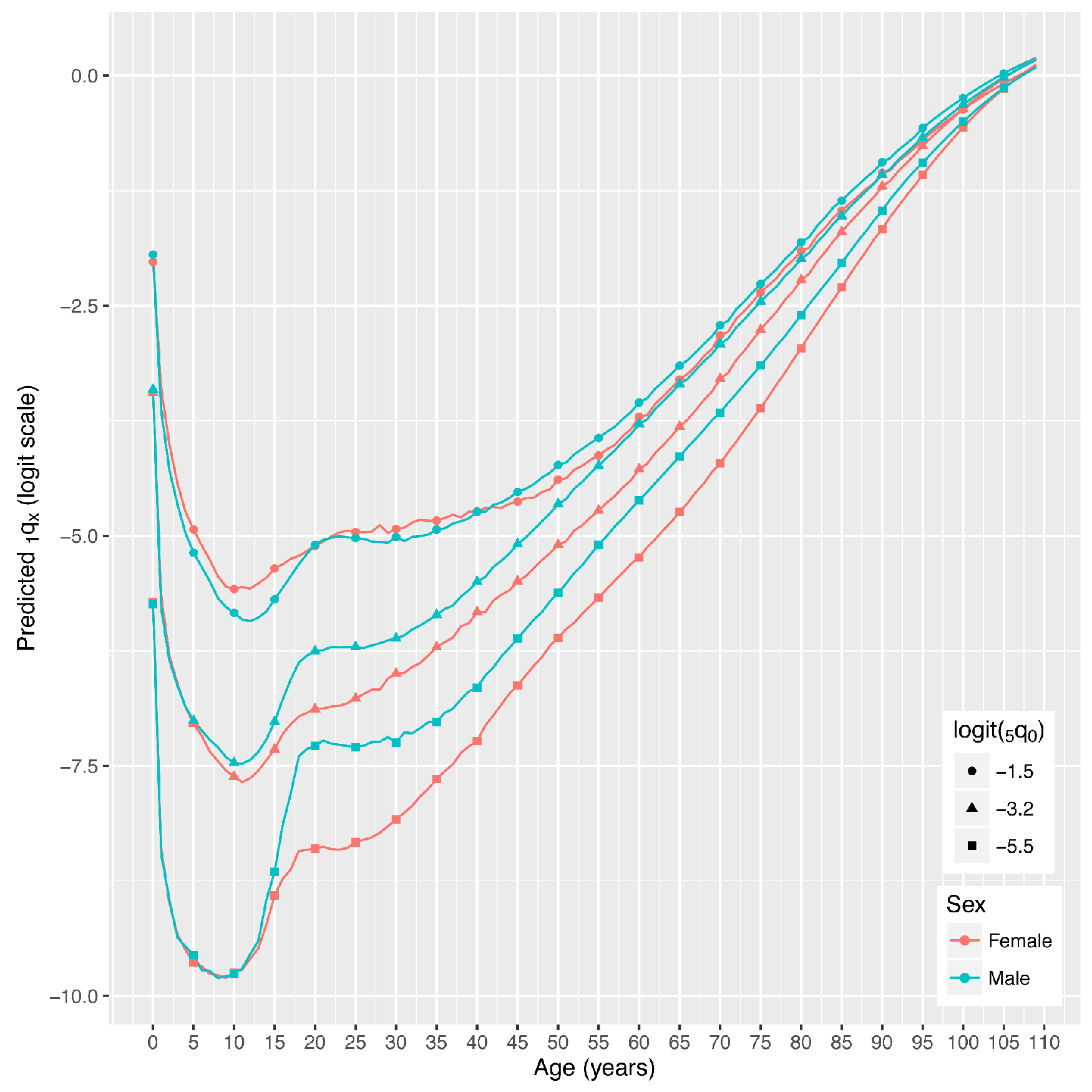} 
   \captionsetup{format=plain,font=normalsize,margin=0cm,justification=justified}
   \caption{\textbf{Predicted $\qox$ at Three Levels of $\qf$.}  As $\qf$ increases the relationship between female and male mortality changes, and female mortality generally exceeds male mortality between ages roughly 10 and 40 for high levels of $\qf$.  It has been verified that this reflects the real change in this relationship embodied in the HMD life tables.}
   \label{fig:preds}
\end{figure}

Finally, Figure \ref{fig:preds} displays predicted $\qox$ from the SVD-Comp using $\qf$ alone for three different levels of $\qf$.

\section{Discussion}

The SVD-Comp model is a simple framework for building mortality models that can be either empirical or analytical.  Its key advantages are
\begin{enumerate*}[label=\arabic*)]
\item a simple linear structure that does not need to be changed to use the model in a variety of ways;
\item a general `interface' through which arbitrary parameters can affect the age pattern of mortality, the weights in Equation \ref{eq:colReconU5mr}; 
\item an ability to handle arbitrary age groups without having to alter the fundamental structure of the model, including very short, like the one-year age groups used here; and finally
\item through its structure, an inherent constraint that ensures that mortality at each age is related to mortality at each other age according to the age patterns reflected in each of the components
\end{enumerate*}.  Along with these, it also satisfies the combined list of desired characteristics for a mortality model enumerated in the introduction.  

This approach is general and allows all-age (in arbitrarily fine age groups) mortality schedules to be predicted from any covariates that are related to age-specific mortality.  This general relationship is quantified in the models that relate the weights in Equation \ref{eq:colReconU5mr} to the covariates.  Allowing this is the fact that the relationship of each age to all others is maintained through the constant components derived from the SVD, and those intra-age relationships are affected all together through the weights on the components.  This constrains the intra-age relationships and relates them to the covariates in a simple, flexible way.  

When the weights are modeled as functions of child mortality and calibrated using the relationship between the empirical weights ($v_{z \ell i}$ in Equation \ref{eq:colRecon}) and child mortality in the HMD, the model serves the same purpose as the Log-Quad \citep{wilmoth2012flexible} model, and it performs slightly better in a direct comparison, while having the advantage of producing mortality schedules by single year of age.  The cross validation results clearly demonstrate that the calibration to the HMD is robust with respect to exactly which and how many mortality schedules are used.  Finally, the SVD-Comp model uses twelve regression models (eight in Equation. \ref{eqn:vsByMx}, two in Equation \ref{eqn:amCm}, and two in Equation \ref{eqn:q0Cm}) to capture the relationship between child mortality and mortality at other ages in the HMD.  In contrast the Log-Quad uses one log-quadratic model of the general form $\log(\mfx) \sim \log(\qf) + \log(\qf)^2$ for each five-year age group and another to refine the prediction of $\qoz$, or at least twenty-two regression models in total.  In addition to a nearly twofold increase in the complexity of the overall model, as measured by the number of submodels required, there is nothing in the overall Log-Quad model to directly constrain the relationship of mortality at one age to another except for the quadratic form of the relationship between mortality at each age and $\qf$.

Together with our earlier work on an HIV-calibrated version of SVD-Comp \citep{sharrow2014modeling}, this demonstration suggests that it is reasonable to expect that SVD-Comp could be calibrated in a variety of additional ways to produce useful models that relate age-specific mortality to, for example, life expectancy at birth (or some other age), GDP, geographic region, time period, epidemiological indicators \citep[as in ][]{sharrow2014modeling}, a combination of any of these, or something else.  Moreover, subtle effects on the age structure of mortality such as the `rotation' in age-specific mortality identified by \cite{li2011} could be incorporated by adding the necessary elements to the models for the weights.  The same approach could be applied to develop models for the difference between underlying age-specific mortality and age-specific mortality affected by specific shocks such as natural disasters, conflict or epidemic disease such as HIV.  It is even possible to refine the Wilmoth/Lee-Carter model in Equation \ref{eq:leeCarter} by adding more components to the SVD-derived $\mbf{b}_xk_t$ term so that the enhanced model could represent a wide range of age patterns instead of the constant age pattern included in the existing formulation.  This would add more parameters to the model, but the payoff might be sufficient to make that worthwhile.  Going further, the entire Wilmoth/Lee-Carter model could be replaced by the SVD-Comp model which would give it the ability to model changing levels and age patterns of mortality independently and generally be far more flexible.

Finally, the general SVD-Comp model in Equation \ref{eq:colReconU5mr} can be used in another way to interpolate or smooth incomplete or noisy age schedules by simply using OLS regression of the incomplete mortality schedule against the corresponding elements of the first few components $s_{zi}\mbf{u}_{zi}$ with the constant constrained to be zero, and then predicting the full mortality schedule from all elements of the components and the coefficients estimated by the regression.   Bayesian estimation can also be used to estimate the weights and their uncertainty, similar to \cite{sharrow2010age}.

An \textsf{R} package \citep{RCore} implementing the HMD child or child/adult mortality-calibrated version of SVD-Comp presented above is available on request and will be available as a fully open source and free to download `\textsf{R} package' on the Comprehensive \textsf{R} Archive Network (CRAN) \citep{CRAN-url}when this article is published.

\section{Acknowledgements}
This work was supported in part by grant R01 HD054511 from the Eunice Kennedy Shriver National Institute of Child Health and Human Development (NICHD).  The funder had no part in the design, execution, or interpretation of the work.  Tables are formatted using the LaTeX package `stargazer' \citep{stargazer}.

\newpage
\bibliography{compModRef-childMx}

\begin{thebibliography}{}

\bibitem[\protect\citeauthoryear{Alexander, Zagheni, and Barbieri}{Alexander
  et~al.}{2016}]{alexander2016flexible}
Alexander, M., E.~Zagheni, and M.~Barbieri (2016).
\newblock A flexible bayesian model for estimating subnational mortality.
\newblock {\em arXiv preprint arXiv:1607.03534\/}.

\bibitem[\protect\citeauthoryear{Bell}{Bell}{1997}]{bell1997comparing}
Bell, W.~R. (1997).
\newblock Comparing and assessing time series methods for forecasting
  age-specific fertility and mortality rates.
\newblock {\em Journal of Official Statistics\/}~{\em 13}.

\bibitem[\protect\citeauthoryear{Bourgeois-Pichat}{Bourgeois-Pichat}{1962}]{bourgeois1962factor}
Bourgeois-Pichat, J. (1962).
\newblock Factor analysis and sex-age-specific death rates: a contribution to
  the study of the dimensions of mortality.
\newblock {\em United Nations Population Bulletin\/}~(6), 147--201.

\bibitem[\protect\citeauthoryear{Bourgeois-Pichat}{Bourgeois-Pichat}{1990}]{bourgeois1990application}
Bourgeois-Pichat, J. (1990).
\newblock Application de l'analyse factorielle {\`a} l'{\'e}tude de la
  mortalit{\'e}.
\newblock {\em Population (french edition)\/}~{\em 45\/}(4-5), 773--802.

\bibitem[\protect\citeauthoryear{Bozik and Bell}{Bozik and
  Bell}{1987}]{bozik1987forecasting}
Bozik, J.~E. and W.~R. Bell (1987).
\newblock Forecasting age specific fertility using principal components.
\newblock In {\em Proceedings of the American Statistical Association, Social
  Statistics Section}, Volume 396, pp.\  401.

\bibitem[\protect\citeauthoryear{Brass}{Brass}{1971}]{brass1971scale}
Brass, W. (1971).
\newblock On the scale of mortality.
\newblock In W.~Brass (Ed.), {\em Biological Aspects of Demography}, pp.\
  69--110. Taylor and Francis: London, UK.

\bibitem[\protect\citeauthoryear{Carter and Lee}{Carter and
  Lee}{1986}]{carter1986joint}
Carter, L.~R. and R.~D. Lee (1986).
\newblock Joint forecasts of us marital fertility, nuptiality, births, and
  marriages using time series models.
\newblock {\em Journal of the American Statistical Association\/}~{\em
  81\/}(396), 902--911.

\bibitem[\protect\citeauthoryear{Clark}{Clark}{2001}]{clarkPhD}
Clark, S.~J. (2001).
\newblock {\em An Investigation into the Impact of HIV on Population Dynamics
  in Africa}.
\newblock Ph.\ D. thesis, University of Pennsylvania.

\bibitem[\protect\citeauthoryear{Clark}{Clark}{2015}]{clark2015singular}
Clark, S.~J. (2015).
\newblock A singular value decomposition-based factorization and parsimonious
  component model of demographic quantities correlated by age: Predicting
  complete demographic age schedules with few parameters.
\newblock {\em arXiv preprint arXiv:1504.02057\/}.

\bibitem[\protect\citeauthoryear{Clark, Jasseh, Punpuing, Zulu, Bawah, and
  Sankoh}{Clark et~al.}{2009}]{clark2009IndMltPaa}
Clark, S.~J., M.~Jasseh, S.~Punpuing, E.~Zulu, A.~Bawah, and O.~Sankoh (2009,
  May).
\newblock Indepth model life tables 2.0.
\newblock In {\em Annual Conference of the Population Association of America}.
  Population Association of America (PAA).

\bibitem[\protect\citeauthoryear{Clark and Sharrow}{Clark and
  Sharrow}{2011a}]{clark2011ContMltPaa}
Clark, S.~J. and D.~J. Sharrow (2011a, April).
\newblock Contemporary model life tables for developed countries -- an
  application of model-based clustering.
\newblock In {\em Annual Conference of the Population Association of America}.
  Population Association of America (PAA).

\bibitem[\protect\citeauthoryear{Clark and Sharrow}{Clark and
  Sharrow}{2011b}]{clarkSharrow2011}
Clark, S.~J. and D.~J. Sharrow (2011b).
\newblock Contemporary model life tables for developed countries: An
  application of model-based clustering.
\newblock {\em Center for Statistics and the Social Sciences (CSSS) Working
  Paper Series\/}~(107).

\bibitem[\protect\citeauthoryear{Coale and Demeny}{Coale and
  Demeny}{1966}]{coale1966}
Coale, A.~J. and P.~Demeny (1966).
\newblock {\em Regional Model Life Tables and Stable Populations}.
\newblock Princeton University Press.

\bibitem[\protect\citeauthoryear{Coale and Trussell}{Coale and
  Trussell}{1974}]{coale1974model}
Coale, A.~J. and T.~J. Trussell (1974).
\newblock Model fertility schedules: variations in the age structure of
  childbearing in human populations.
\newblock {\em Population Index\/}~(1974), 185--258.

\bibitem[\protect\citeauthoryear{Fosdick and Hoff}{Fosdick and
  Hoff}{2012}]{fosdick2012separable}
Fosdick, B.~K. and P.~D. Hoff (2012).
\newblock Separable factor analysis with applications to mortality data.
\newblock {\em arXiv preprint arXiv:1211.3813\/}.

\bibitem[\protect\citeauthoryear{Golub, Hoffman, and Stewart}{Golub
  et~al.}{1987}]{golub1987generalization}
Golub, G.~H., A.~Hoffman, and G.~W. Stewart (1987).
\newblock A generalization of the eckart-young-mirsky matrix approximation
  theorem.
\newblock {\em Linear Algebra and Its Applications\/}~{\em 88}, 317--327.

\bibitem[\protect\citeauthoryear{Gompertz}{Gompertz}{1825}]{gompertz1825nature}
Gompertz, B. (1825).
\newblock On the nature of the function expressive of the law of human
  mortality, and on a new mode of determining the value of life contingencies.
\newblock {\em Philosophical transactions of the Royal Society of
  London\/}~{\em 115}, 513--583.

\bibitem[\protect\citeauthoryear{Good}{Good}{1969}]{good1969some}
Good, I.~J. (1969).
\newblock Some applications of the singular decomposition of a matrix.
\newblock {\em Technometrics\/}~{\em 11\/}(4), 823--831.

\bibitem[\protect\citeauthoryear{Heligman and Pollard}{Heligman and
  Pollard}{1980}]{heligman1980}
Heligman, L. and J.~H. Pollard (1980).
\newblock The age pattern of mortality.
\newblock {\em Journal of the Institute of Actuaries\/}~{\em 107\/}(434),
  49--80.

\bibitem[\protect\citeauthoryear{Hlavac}{Hlavac}{2015}]{stargazer}
Hlavac, M. (2015).
\newblock {\em stargazer: Well-Formatted Regression and Summary Statistics
  Tables}.
\newblock Cambridge, USA: Harvard University.
\newblock R package version 5.2.

\bibitem[\protect\citeauthoryear{{INDEPTH Network}}{{INDEPTH
  Network}}{2002}]{indepthMLT2002}
{INDEPTH Network} (2002).
\newblock {\em INDEPTH Mortality Patterns for Africa}, Volume~1 of {\em
  Population and Health in Developing Countries}, Chapter~7, pp.\  83--128.
\newblock Ottawa: IDRC Press.

\bibitem[\protect\citeauthoryear{Ledermann}{Ledermann}{1969}]{ledermann1969nouvelles}
Ledermann, S. (1969).
\newblock Nouvelles tables-types de mortalit{\'e}.
\newblock Number~53 in INED Traveaux et Documents. Paris: Presses
  Universitaires de France.

\bibitem[\protect\citeauthoryear{Ledermann and Breas}{Ledermann and
  Breas}{1959}]{ledermann1959dimensions}
Ledermann, S. and J.~Breas (1959).
\newblock Les dimensions de la mortalit{\'e}.
\newblock {\em Population (french edition)\/}, 637--682.

\bibitem[\protect\citeauthoryear{Lee}{Lee}{1993}]{lee1993modeling}
Lee, R.~D. (1993).
\newblock Modeling and forecasting the time series of {US} fertility: Age
  distribution, range, and ultimate level.
\newblock {\em International Journal of Forecasting\/}~{\em 9\/}(2), 187--202.

\bibitem[\protect\citeauthoryear{Lee and Carter}{Lee and
  Carter}{1992}]{lee1992modeling}
Lee, R.~D. and L.~R. Carter (1992).
\newblock Modeling and forecasting {US} mortality.
\newblock {\em Journal of the American statistical association\/}~{\em
  87\/}(419), 659--671.

\bibitem[\protect\citeauthoryear{Li}{Li}{2015}]{li2015wppLt}
Li, N. (2015).
\newblock Estimating life tables for developing countries.
\newblock Technical Report 2014/4, United Nations Department of Economic and
  Social Affairs Population Division,
  \url{http://www.un.org/en/development/desa/population/publications/pdf/technical/TP2014-4.pdf}.

\bibitem[\protect\citeauthoryear{Li and Gerland}{Li and Gerland}{2011}]{li2011}
Li, N. and P.~Gerland (2011).
\newblock Modifying the {Lee-Carter} method to project mortality changes up to
  2100.
\newblock Paper presented at the 2011 Annual Meeting of the Population
  Association of America (PAA), Washington, D.C., March 31-April 2.

\bibitem[\protect\citeauthoryear{Li and Anderson}{Li and
  Anderson}{2009}]{li2009vitality}
Li, T. and J.~J. Anderson (2009).
\newblock The vitality model: A way to understand population survival and
  demographic heterogeneity.
\newblock {\em Theoretical Population Biology\/}~{\em 76\/}(2), 118--131.

\bibitem[\protect\citeauthoryear{Makeham}{Makeham}{1860}]{makeham1860law}
Makeham, W.~M. (1860).
\newblock On the law of mortality and the construction of annuity tables.
\newblock {\em The Assurance Magazine, and Journal of the Institute of
  Actuaries\/}~{\em 8\/}(6), 301--310.

\bibitem[\protect\citeauthoryear{Murray, Ferguson, Lopez, Guillot, Salomon, and
  Ahmad}{Murray et~al.}{2003}]{murray2003}
Murray, C.~J., B.~D. Ferguson, A.~D. Lopez, M.~Guillot, J.~A. Salomon, and
  O.~Ahmad (2003).
\newblock Modified logit life table system: principles, empirical validation,
  and application.
\newblock {\em Population Studies\/}~{\em 57\/}(2), 165--182.

\bibitem[\protect\citeauthoryear{{\textsf{R} Core Team}}{{\textsf{R} Core
  Team}}{2016}]{RCore}
{\textsf{R} Core Team} (2016).
\newblock {\em R: A Language and Environment for Statistical Computing}.
\newblock Vienna, Austria: \textsf{R} Foundation for Statistical Computing.

\bibitem[\protect\citeauthoryear{{\textsf{R} Foundation for Statistical
  Computing}}{{\textsf{R} Foundation for Statistical
  Computing}}{2016a}]{CRAN-url}
{\textsf{R} Foundation for Statistical Computing} (2016a).
\newblock {\em The Comprehensive \textsf{R} Archive Network - CRAN}.
\newblock \url{https://cran.r-project.org}.

\bibitem[\protect\citeauthoryear{{\textsf{R} Foundation for Statistical
  Computing}}{{\textsf{R} Foundation for Statistical Computing}}{2016b}]{R-url}
{\textsf{R} Foundation for Statistical Computing} (2016b).
\newblock {\em The \textsf{R} Project for Statistical Computing}.
\newblock \url{http://www.r-project.org}.

\bibitem[\protect\citeauthoryear{Sharrow, Clark, Collinson, Kahn, and
  Tollman}{Sharrow et~al.}{2010}]{sharrow2010age}
Sharrow, D.~J., S.~J. Clark, M.~A. Collinson, K.~Kahn, and S.~M. Tollman
  (2010).
\newblock The age-pattern of increases in mortality affected by hiv: Bayesian
  fit of the heligman-pollard model to data from the agincourt hdss field site
  in rural northeast south africa.
\newblock {\em University of Washington\/}.

\bibitem[\protect\citeauthoryear{Sharrow, Clark, and Raftery}{Sharrow
  et~al.}{2014}]{sharrow2014modeling}
Sharrow, D.~J., S.~J. Clark, and A.~E. Raftery (2014).
\newblock Modeling age-specific mortality for countries with generalized hiv
  epidemics.
\newblock {\em PloS ONE\/}~{\em 9\/}(5), e96447.

\bibitem[\protect\citeauthoryear{Stewart}{Stewart}{1993}]{stewart1993early}
Stewart, G.~W. (1993).
\newblock On the early history of the singular value decomposition.
\newblock {\em SIAM review\/}~{\em 35\/}(4), 551--566.

\bibitem[\protect\citeauthoryear{Strang}{Strang}{2009}]{strang2009introduction}
Strang, G. (2009).
\newblock {\em Introduction to Linear Algebra 4e}.
\newblock Wellesley-Cambridge Press.

\bibitem[\protect\citeauthoryear{{United Nations, Department of Economic and
  Social Affairs, Population Division}}{{United Nations, Department of Economic
  and Social Affairs, Population Division}}{1955}]{united1955age}
{United Nations, Department of Economic and Social Affairs, Population
  Division} (1955).
\newblock {\em Age and Sex Patterns of Mortality: Model Life-tables for
  Under-developed Countries}.
\newblock New York: United Nations Department of International Economic and
  Social Affairs Population Division.

\bibitem[\protect\citeauthoryear{{United Nations, Department of Economic and
  Social Affairs, Population Division}}{{United Nations, Department of Economic
  and Social Affairs, Population Division}}{1982}]{united1982model}
{United Nations, Department of Economic and Social Affairs, Population
  Division} (1982).
\newblock {\em Model life tables for developing countries}.
\newblock Number~77. New York: United Nations Department of International
  Economic and Social Affairs Population Division.

\bibitem[\protect\citeauthoryear{{United Nations, Department of Economic and
  Social Affairs, Population Division}}{{United Nations, Department of Economic
  and Social Affairs, Population Division}}{2015a}]{unWPP2015Meta}
{United Nations, Department of Economic and Social Affairs, Population
  Division} (2015a).
\newblock {File 0-2: Latest data sources used to derive estimates for total
  population, fertility, mortality and migration by countries or areas in WPP
  2015 revision: POP/DB/WPP/Rev.2015/F0-2}.
\newblock
  {\url{https://esa.un.org/unpd/wpp/DVD/Files/4\_Other\%20Files/WPP2015\_F02\_METAINFO.XLS}}.

\bibitem[\protect\citeauthoryear{{United Nations, Department of Economic and
  Social Affairs, Population Division}}{{United Nations, Department of Economic
  and Social Affairs, Population Division}}{2015b}]{un2015}
{United Nations, Department of Economic and Social Affairs, Population
  Division} (2015b).
\newblock {\em World Population Prospects: the 2015 Revision}.
\newblock New York: United Nations.

\bibitem[\protect\citeauthoryear{{United Nations, Department of Economic and
  Social Affairs, Population Division}}{{United Nations, Department of Economic
  and Social Affairs, Population Division}}{2015c}]{unWPPMethods2015}
{United Nations, Department of Economic and Social Affairs, Population
  Division} (2015c).
\newblock {\em World Population Prospects: The 2015 Revision, Methodology of
  the United Nations Population Estimates and Projections}.
\newblock Working paper No. ESA/P/WP.242.

\bibitem[\protect\citeauthoryear{{University of California, Berkeley} and {Max
  Planck Institute for Demographic Research}}{{University of California,
  Berkeley} and {Max Planck Institute for Demographic
  Research}}{2016}]{hmd2016}
{University of California, Berkeley} and {Max Planck Institute for Demographic
  Research} (Downloaded November 2016).
\newblock {\em Human Mortality Database}.
\newblock \url{http://www.mortality.org} or \url{http://www.humanmortality.de}.

\bibitem[\protect\citeauthoryear{Wang, Dwyer-Lindgren, Lofgren, Rajaratnam,
  Marcus, Levin-Rector, Levitz, Lopez, and Murray}{Wang
  et~al.}{2013}]{Wang2013}
Wang, H., L.~Dwyer-Lindgren, K.~T. Lofgren, J.~K. Rajaratnam, J.~R. Marcus,
  A.~Levin-Rector, C.~E. Levitz, A.~D. Lopez, and C.~J.~L. Murray (2013).
\newblock Age-specific and sex-specific mortality in 187 countries, 1970--2010:
  a systematic analysis for the global burden of disease study 2010.
\newblock {\em The Lancet\/}~{\em 380\/}(9859), 2071--2094.

\bibitem[\protect\citeauthoryear{Wilmoth, Vallin, and Caselli}{Wilmoth
  et~al.}{1989}]{wilmoth1989quand}
Wilmoth, J., J.~Vallin, and G.~Caselli (1989).
\newblock Quand certaines g{\'e}n{\'e}rations ont une mortalit{\'e}
  diff{\'e}rente de celle que l'on pourrait attendre.
\newblock {\em Population\/}~{\em 44\/}(2), 335--376.

\bibitem[\protect\citeauthoryear{Wilmoth, Zureick, Canudas-Romo, Inoue, and
  Sawyer}{Wilmoth et~al.}{2012}]{wilmoth2012flexible}
Wilmoth, J., S.~Zureick, V.~Canudas-Romo, M.~Inoue, and C.~Sawyer (2012).
\newblock A flexible two-dimensional mortality model for use in indirect
  estimation.
\newblock {\em Population studies\/}~{\em 66\/}(1), 1--28.

\bibitem[\protect\citeauthoryear{Wilmoth}{Wilmoth}{1988}]{wilmoth1988Phd}
Wilmoth, J.~R. (1988).
\newblock {\em On the Statistical Analysis of Large Arrays of Demographic
  Rates}.
\newblock Ph.\ D. thesis, Department of Statistics, Princeton University.

\bibitem[\protect\citeauthoryear{Wilmoth}{Wilmoth}{1990}]{wilmoth1990variation}
Wilmoth, J.~R. (1990).
\newblock Variation in vital rates by age, period, and cohort.
\newblock {\em Sociological Methodology\/}~{\em 20}, 295--335.

\bibitem[\protect\citeauthoryear{Wilmoth and Caselli}{Wilmoth and
  Caselli}{1987}]{wilmoth1987simple}
Wilmoth, J.~R. and G.~Caselli (1987).
\newblock A simple model for the statistical analysis of large arrays of
  mortality data: rectangular vs. diagonal structure.
\newblock {\em IIASA Working Paper\/}~(WP-87-058).

\bibitem[\protect\citeauthoryear{Zaba}{Zaba}{1979}]{zaba1979four}
Zaba, B. (1979).
\newblock The four-parameter logit life table system.
\newblock {\em Population Studies\/}~{\em 33\/}(1), 79--100.

\end{thebibliography}
\bibliographystyle{chicago}

\newpage
\begin{appendices}
\counterwithin{figure}{section} 
\counterwithin{table}{section}
\counterwithin{equation}{section}

\section{Estimated Regression Coefficients} \label{app:regs}

\begin{table}[!htbp] \centering 
  \caption{Female RSV Models: $v_{\ell i} = f_{i}(\qf_{\, \ell},\qff_{\, \ell})$} 
  \label{tab:appA:femaleRSVMods} 
\begin{tabular}{@{\extracolsep{5pt}}lcccc} 
\toprule
 & \multicolumn{4}{c}{\textit{Dependent variable:}} \\ 
\cline{2-5} 
\\[-1.8ex] & $\mbf{v}_1$ & $\mbf{v}_2$ & $\mbf{v}_3$ & $\mbf{v}_4$ \\ 
\\[-1.8ex] & (1) & (2) & (3) & (4)\\ 
\midrule
 $\qf$ & 0.017$^{***}$ & 0.521$^{***}$ & $-$0.814$^{***}$ & 1.901$^{***}$ \\ 
  & (0.001) & (0.045) & (0.101) & (0.100) \\ 
  & & & & \\ 
 $\logit(\qf)$ & $-$0.005$^{***}$ & $-$0.162$^{***}$ & 0.211$^{***}$ & $-$0.525$^{***}$ \\ 
  & (0.0004) & (0.013) & (0.030) & (0.030) \\ 
  & & & & \\ 
 $\logit(\qf)^2$ & $-$0.001$^{***}$ & $-$0.030$^{***}$ & 0.025$^{***}$ & $-$0.104$^{***}$ \\ 
  & (0.0001) & (0.003) & (0.006) & (0.006) \\ 
  & & & & \\ 
 $\logit(\qf)^3$ & $-$0.0001$^{***}$ & $-$0.002$^{***}$ & 0.002$^{***}$ & $-$0.007$^{***}$ \\ 
  & (0.00001) & (0.0002) & (0.0004) & (0.0004) \\ 
  & & & & \\ 
 $\qff$ & $-$0.003$^{***}$ & $-$0.005 & 0.074$^{***}$ & $-$0.055$^{***}$ \\ 
  & (0.0001) & (0.005) & (0.010) & (0.010) \\ 
  & & & & \\ 
 $\logit(\qff)^2$ & 0.0004$^{***}$ & 0.013$^{***}$ & $-$0.023$^{***}$ & 0.014$^{***}$ \\ 
  & (0.00002) & (0.001) & (0.002) & (0.002) \\ 
  & & & & \\ 
 $\logit(\qff)^3$ & $-$0.00002$^{***}$ & 0.002$^{***}$ & 0.003$^{***}$ & 0.002$^{***}$ \\ 
  & (0.00001) & (0.0002) & (0.0004) & (0.0004) \\ 
  & & & & \\ 
 $\qf \times \qff$ & $-$0.0004$^{***}$ & $-$0.007$^{***}$ & 0.043$^{***}$ & $-$0.004$^{**}$ \\ 
  & (0.00002) & (0.001) & (0.002) & (0.002) \\ 
  & & & & \\ 
 Constant & 0.006$^{***}$ & $-$0.294$^{***}$ & 0.359$^{***}$ & $-$0.912$^{***}$ \\ 
  & (0.001) & (0.023) & (0.051) & (0.051) \\ 
\midrule
Observations & 4,486 & 4,486 & 4,486 & 4,486 \\ 
R$^{2}$ & 0.966 & 0.860 & 0.308 & 0.319 \\ 
Adjusted R$^{2}$ & 0.966 & 0.860 & 0.306 & 0.318 \\ 
Residual Std. Error (df = 4477) & 0.0002 & 0.006 & 0.012 & 0.012 \\ 
F Statistic (df = 8; 4477) & 16,031.850$^{***}$ & 3,433.656$^{***}$ & 248.516$^{***}$ & 262.679$^{***}$ \\ 
\bottomrule
\multicolumn{5}{c}{$^{*}$p$<$0.1; $^{**}$p$<$0.05; $^{***}$p$<$0.01} \\ 
\end{tabular} 
\end{table}

\begin{table}[!htbp] \centering 
  \caption{Male RSV Models: $v_{\ell i} = f_{i}(\qf_{\, \ell},\qff_{\, \ell})$} 
  \label{tab:appA:maleRSVMods} 
\begin{tabular}{@{\extracolsep{5pt}}lcccc} 
\toprule
 & \multicolumn{4}{c}{\textit{Dependent variable:}} \\ 
\cline{2-5} 
\\[-1.8ex] & $\mbf{v}_1$ & $\mbf{v}_2$ & $\mbf{v}_3$ & $\mbf{v}_4$ \\ 
\\[-1.8ex] & (1) & (2) & (3) & (4)\\ 
\midrule
 $\qf$ & 0.012$^{***}$ & 0.320$^{***}$ & 0.532$^{***}$ & $-$2.081$^{***}$ \\ 
  & (0.001) & (0.045) & (0.084) & (0.104) \\ 
  & & & & \\ 
 $\logit(\qf)$ & $-$0.004$^{***}$ & $-$0.110$^{***}$ & $-$0.145$^{***}$ & 0.588$^{***}$ \\ 
  & (0.0003) & (0.014) & (0.025) & (0.031) \\ 
  & & & & \\ 
 $\logit(\qf)^2$ & $-$0.001$^{***}$ & $-$0.021$^{***}$ & $-$0.031$^{***}$ & 0.112$^{***}$ \\ 
  & (0.0001) & (0.003) & (0.005) & (0.006) \\ 
  & & & & \\ 
 $\logit(\qf)^3$ & $-$0.0001$^{***}$ & $-$0.002$^{***}$ & $-$0.002$^{***}$ & 0.007$^{***}$ \\ 
  & (0.00000) & (0.0002) & (0.0004) & (0.0005) \\ 
  & & & & \\ 
 $\qff$ & $-$0.002$^{***}$ & $-$0.006$^{**}$ & $-$0.109$^{***}$ & 0.066$^{***}$ \\ 
  & (0.0001) & (0.003) & (0.006) & (0.007) \\ 
  & & & & \\ 
 $\logit(\qff)^2$ & 0.0001$^{***}$ & 0.002$^{***}$ & 0.002$^{***}$ & 0.005$^{***}$ \\ 
  & (0.00001) & (0.0004) & (0.001) & (0.001) \\ 
  & & & & \\ 
 $\logit(\qff)^3$ & $-$0.00001$^{***}$ & 0.001$^{***}$ & 0.001$^{***}$ & 0.001$^{**}$ \\ 
  & (0.00000) & (0.0001) & (0.0003) & (0.0003) \\ 
  & & & & \\ 
 $\qf \times \qff$ & $-$0.00004$^{***}$ & $-$0.0004 & 0.004$^{***}$ & 0.004$^{***}$ \\ 
  & (0.00001) & (0.0004) & (0.001) & (0.001) \\ 
  & & & & \\ 
 Constant & 0.009$^{***}$ & $-$0.195$^{***}$ & $-$0.214$^{***}$ & 1.009$^{***}$ \\ 
  & (0.0005) & (0.023) & (0.043) & (0.053) \\ 
\midrule
Observations & 4,486 & 4,486 & 4,486 & 4,486 \\ 
R$^{2}$ & 0.974 & 0.874 & 0.562 & 0.329 \\ 
Adjusted R$^{2}$ & 0.974 & 0.874 & 0.562 & 0.328 \\ 
Residual Std. Error (df = 4477) & 0.0001 & 0.005 & 0.010 & 0.012 \\ 
F Statistic (df = 8; 4477) & 21,228.310$^{***}$ & 3,892.337$^{***}$ & 719.216$^{***}$ & 274.413$^{***}$ \\ 
\bottomrule
\multicolumn{5}{c}{$^{*}$p$<$0.1; $^{**}$p$<$0.05; $^{***}$p$<$0.01} \\ 
\end{tabular} 
\end{table}

\begin{table}[!htbp] \centering 
  \captionsetup{format=hang,font=normalsize,margin=3.2cm,justification=justified}
  \caption{Adult Mortality Models: \\ $\logit(\qff)_{z \ell} = f(\qf_{\, z \ell})$} 
  \label{tab:appA:adultMxMod} 
\begin{tabular}{@{\extracolsep{5pt}}lcc} 
\toprule
 & \multicolumn{2}{c}{\textit{Dependent variable:}} \\ 
\cline{2-3} 
\\[-1.8ex] & \multicolumn{2}{c}{logit$(\qff)$} \\ 
 & female & male \\ 
\midrule
 $\qf$ & $-$11.468$^{***}$ & $-$0.398 \\ 
  & (1.805) & (2.701) \\ 
  & & \\ 
 $\logit(\qf)$ & 4.208$^{***}$ & 1.359$^{*}$ \\ 
  & (0.538) & (0.814) \\ 
  & & \\ 
 $\logit(\qf)^2$ & 0.735$^{***}$ & 0.313$^{*}$ \\ 
  & (0.109) & (0.167) \\ 
  & & \\ 
 $\logit(\qf)^3$ & 0.049$^{***}$ & 0.031$^{***}$ \\ 
  & (0.008) & (0.012) \\ 
  & & \\ 
 Constant & 6.264$^{***}$ & 0.976 \\ 
  & (0.919) & (1.382) \\ 
\midrule
Observations & 4,486 & 4,486 \\ 
R$^{2}$ & 0.932 & 0.789 \\ 
F Statistic (df = 4; 4481) & 15,470.360$^{***}$ & 4,199.566$^{***}$ \\ 
\bottomrule
\multicolumn{3}{c}{$^{*}$p$<$0.1; $^{**}$p$<$0.05; $^{***}$p$<$0.01} \\ 
\end{tabular} 
\end{table}

\begin{table}[!htbp] \centering 
  \caption{Infant MortalityModels: $\logit(\qoz)_{z \ell} = f(\qf_{\, z \ell})$}
  \label{tab:appA:infantMxMod} 
\begin{tabular}{@{\extracolsep{5pt}}lcc} 
\toprule
 & \multicolumn{2}{c}{\textit{Dependent variable:}} \\ 
\cline{2-3} 
\\[-1.8ex] & \multicolumn{2}{c}{logit$(\qoz)$} \\ 
 & female & male \\ 
\\[-1.8ex] & (1) & (2)\\ 
\midrule
 $\logit(\qf)$ & 0.658$^{***}$ & 0.686$^{***}$ \\ 
  & (0.005) & (0.004) \\ 
  & & \\ 
 $\logit(\qf)^2$ & $-$0.038$^{***}$ & $-$0.038$^{***}$ \\ 
  & (0.001) & (0.001) \\ 
  & & \\ 
 Constant & $-$0.951$^{***}$ & $-$0.830$^{***}$ \\ 
  & (0.006) & (0.006) \\ 
\midrule
Observations & 4,486 & 4,486 \\ 
R$^{2}$ & 0.995 & 0.996 \\ 
F Statistic (df = 2; 4483) & 485,867.000$^{***}$ & 543,813.000$^{***}$ \\ 
\bottomrule
\multicolumn{3}{c}{$^{*}$p$<$0.1; $^{**}$p$<$0.05; $^{***}$p$<$0.01} \\ 
\end{tabular} 
\end{table}

\newpage
\section{SVD Relationship Algebra} \label{app:svd}

Below I rearrange the basic SVD relationship to derive useful additional relationships. %
\begin{align}
\mbf{X}
&=
\mbf{USV}^\text{T} 
\\
\left[ 
	\begin{matrix}
	| & & | \\
	\mbf{x}_{1} & \ldots  & \mbf{x}_{L}  \\
	| & & | \\
	\end{matrix}
\right] 
&=
\left[ 
	\begin{matrix}
	| & & | \\
	\mbf{u}_{1} & \ldots  & \mbf{u}_{\rho}  \\
	| & & |  \\
	\end{matrix}
\right]  
\left[ 
	\begin{matrix}
		s_{1} & \ldots & 0  \\
		\vdots & \ddots & \vdots \\
		0 & \ldots & s_{\rho}  \\
	\end{matrix}
\right]
\left[ 
	\begin{matrix}
		\text{---} & \mbf{v}_1 & \text{---} \\
		& \vdots  & \\
		\text{---} & \mbf{v}_\rho & \text{---} \\
	\end{matrix}
\right] 	 
\nonumber \\
&=
\left[ 
	\begin{matrix}
		| & & | \\
	\mbf{u}_{1} & \ldots  & \mbf{u}_{\rho}  \\
		| & & | \\
	\end{matrix}
\right]  
\left[ 
	\begin{matrix}
		\text{---} & s_{1} \mbf{v}_1 & \text{---} \\
		& \vdots & \\
		\text{---} & s_{\rho} \mbf{v}_\rho & \text{---} \\
	\end{matrix}
\right] 	 
\nonumber \\
&=
\left[ 
	\begin{matrix}
		\sum_{i=1}^{\rho} u_{1i} s_{i} v_{1i} & \ldots & \sum_{i=1}^{\rho} u_{1i} s_{i} v_{Li} \\
		\vdots & \ddots  & \vdots \\		
		\sum_{i=1}^{\rho} u_{Ki} s_{i} v_{1i} & \ldots & \sum_{i=1}^{\rho} u_{Ki} s_{i} v_{Li} \\
	\end{matrix}
\right]  
\nonumber \\
&=
\left[ 
	\begin{matrix}
		| & & | \\
		\sum_{i=1}^{\rho} s_{i} v_{1i} \mbf{u}_{i}  & \ldots & \sum_{i=1}^{\rho} s_{i} v_{Li} \mbf{u}_{i}  \\
		| & & | \\
	\end{matrix}
\right] 
\label{eq:appB:genComponents} \\ 
&=
\sum_{i=1}^{\rho} \left[ 
	\begin{matrix}
		| & & | \\
		s_{i} v_{1i} \mbf{u}_{i}  & \ldots  & s_{i} v_{Li} \mbf{u}_{i} \\
		| & & | \\
	\end{matrix}
\right] 
\nonumber \\
&=
\sum_{i=1}^{\rho} \left[ 
	\begin{matrix}
		s_{i} v_{1i} u_{1i} & \ldots & s_{i} v_{Li} u_{1i} \\
		\vdots & \ddots & \vdots \\
		s_{i} v_{1i} u_{Ki}  & \ldots  & s_{i} v_{Li} u_{Ki} \\
	\end{matrix}
\right] 
\nonumber \\
&=
\sum_{i=1}^{\rho}
s_{i} 
\left[ 
	\begin{matrix}
		u_{1i} \\
		\vdots \\
		u_{Ki} \\
	\end{matrix}
\right]  
\left[ 
	\begin{matrix}
		v_{1i} \ldots v_{Li} \\
	\end{matrix}
\right] 
\label{eq:appB:XsumRank-1Detail} \\
\mathbf{X} &=
\sum_{i=1}^{\rho} s_{i} \mbf{u}_{i} \mbf{v}_{i}^\text{T}
\label{eq:appB:XsumRank-1}
\end{align}%

From Equation \ref{eq:appB:genComponents} we have
\begin{align}
\mbf{x}_{\ell} = \sum_{i=1}^{\rho} s_{i} v_{\ell i} \mbf{u}_{i} \ .
\label{eq:appB:cols}
\end{align}%

\end{appendices}

\end{document}